\documentclass[aps,pre,twocolumn,showpacs,superscriptaddress]{revtex4-1}
\usepackage{graphicx}
\usepackage{dcolumn,xcolor,ulem}
\usepackage{amsmath} 
\usepackage{amssymb}
\usepackage{amsfonts}
\usepackage{bm}
\usepackage[latin1]{inputenc}
\usepackage{fancyhdr}
\usepackage{amsmath} 
\usepackage{amssymb,txfonts}

\begin{document}

\title{Avalanche-like fluidization of a non-Brownian particle gel}

\author{Aika Kurokawa}
\email[]{kurokawa@eri.u-tokyo.ac.jp}
\affiliation{Earthquake Research Institute, The University of Tokyo, 1-1-1 Yayoi, Bunkyo-ku, Tokyo, Japan}
\author{Val\'erie Vidal}
\email[]{valerie.vidal@ens-lyon.fr}
\affiliation{Universit${\acute{e}}$ de Lyon, Laboratoire de Physique, $\acute{E}$cole Normale Sup$\acute{e}$rieure de Lyon, CNRS UMR 5672 - 46 All${\acute{e}}$e d'Italie, 69364 Lyon cedex 07, France}
\author{Kei Kurita}
\email[]{kurikuri@eri.u-tokyo.ac.jp}
\affiliation{Earthquake Research Institute, The University of Tokyo, 1-1-1 Yayoi, Bunkyo-ku, Tokyo, Japan}
\author{Thibaut Divoux}
\email[]{divoux@crpp-bordeaux.cnrs.fr}
\affiliation{Universit\'e de Bordeaux, Centre de Recherche Paul Pascal, CNRS UPR~8641, 115 av. Dr. Schweitzer, 33600 Pessac, France}
\author{S\'ebastien Manneville}
\email[]{sebastien.manneville@ens-lyon.fr}
\affiliation{Universit${\acute{e}}$ de Lyon, Laboratoire de Physique, $\acute{E}$cole Normale Sup$\acute{e}$rieure de Lyon, CNRS UMR 5672 - 46 All${\acute{e}}$e d'Italie, 69364 Lyon cedex 07, France}

\date{\today}

\begin{abstract}
We report on the fluidization dynamics of an attractive gel composed of non-Brownian particles made of fused silica colloids. Extensive rheology coupled to ultrasonic velocimetry allows us to characterize the global stress response together with the local dynamics of the gel during shear startup experiments. In practice, after being rejuvenated by a preshear, the gel is left to age during a time $t_w$ before being submitted to a constant shear rate $\dot \gamma$. We investigate in detail the effects of both $t_w$ and $\dot \gamma$ on the fluidization dynamics and build a detailed state diagram of the gel response to shear startup flows. The gel may either display transient shear banding towards complete fluidization, or steady-state shear banding. In the former case, we unravel that the progressive fluidization occurs by successive steps that appear as peaks on the global stress relaxation signal. Flow imaging reveals that the shear band grows up to complete fluidization of the material by sudden avalanche-like events which are distributed heterogeneously along the vorticity direction and correlated to large peaks in the slip velocity at the moving wall. These features are robust over a wide range of values of $t_w$ and $\dot \gamma$, although the very details of the fluidization scenario vary with $\dot \gamma$. Finally, the critical shear rate $\dot \gamma^*$ that separates steady-state shear-banding from steady-state homogeneous flow depends on the width on the shear cell and exhibits a nonlinear dependence with $t_w$. Our work brings about valuable experimental data on transient flows of attractive dispersions, highlighting the subtle interplay between shear, wall slip and aging which modeling constitutes a major challenge that has not been met yet.
\end{abstract}


\maketitle

\section{Introduction}

Yield stress materials denote assemblies of mesoscopic constituents such as colloids, droplets or bubbles that display macroscopic properties intermediate between those of fluid and solid  \cite{Coussot:2015,Bonn:2015}. At rest, or under low stresses, yield stress materials display a solid-like behaviour whereas they flow like liquids above a certain threshold referred to as the yield stress  \cite{Moller:2009a,Ovarlez:2013b,Balmforth:2014}. Their solid-like behaviour at rest originates either from a densely packed microstructure composed of soft objects \cite{Bonnecaze:2010} such as in dense emulsions \cite{Mason:1996a}, jammed microgels \cite{Cloitre:2003a,Seth:2006,Siebenburger:2012b}, etc., or from the existence of weak attractive interactions that bind the constituents together and result in the formation of a sample-spanning network \cite{Lu:2008}. In the latter case, solid-like properties emerge even at low packing fractions. This defines a category referred to as \textit{attractive gels} \cite{Lu:2013} which encompasses various systems such as clay suspensions \cite{Mourchid:1995,Pignon:1997b,Paineau:2011}, carbon black gels \cite{Trappe:2000,Trappe:2001} and colloid-polymer mixtures \cite{Pham:2004}. All sorts of colloids and larger particles of different shapes and attraction potential also comply with this definition \cite{Mock:2007,Reddy:2012}. 

Over the past ten years, the rheological behaviour of attractive gels has proved to be by far one of the most challenging to understand among non-Newtonian fluids. In particular, their mechanical properties at rest are strongly time-dependent: attractive gels display a reversible aging dynamics driven by the weak attractive forces between its constituents \cite{Cipelletti:2000} and that can be reversed by shear. As a result, the gel elastic properties display a slow logarithmic or weak power-law increase with time \cite{Derec:2003,Ovarlez:2008b,Negi:2009b,Koumakis:2011}, up to complete demixing of the system, which is historically referred to as syneresis \cite{Scherer:1989}. To make matter worse, particles are often denser than the surrounding fluid which fosters syneresis and may trigger the collapse of the gel \cite{Manley:2005,Buscall:2009,Brambilla:2011,Teece:2011,Barlett:2012}. Such density mismatch is further suspected to influence the behaviour of these systems under flow, although clear experimental evidence is still lacking \cite{Buscall:1983}. Nonetheless, it is now well established that the behaviour of attractive gels under external shear involves heterogeneous flows that are highly sensitive to preshear history, boundary conditions, and/or finite size effects. For instance, one can emphasize the case of Laponite suspensions which steady-state flow properties were shown to be influenced by the nature of the boundaries: under external shear, smooth walls lead to the complete fluidization of the gel and to linear velocity profiles, while rough boundary conditions allow the growth of a steady shear band \cite{Gibaud:2008,Gibaud:2009}. Moreover, as an illustration of the impact of confinement on flows of attractive gels, one can mention the spectacular shear-induced structuration observed at moderate shear rates and reported in silica suspensions \cite{DeGroot:1994}, gels of anisotropic particles \cite{Navarrete:1996}, attractive emulsions \cite{Montesi:2004} and carbon black gels \cite{Osuji:2008,Negi:2009,Grenard:2011}. In such cases, the gel rearranges to form a striped pattern of logrolling flocs aligned along the vorticity direction, which origin and formation mechanism are still highly debated \cite{Vermant:2005}.

Beyond the effects of bounding walls and confinement, attractive interactions alone are also responsible for long-lasting transients under external shear. On the one hand, experiments performed under constant \cite{Gopalakrishnan:2007,Gibaud:2010,Sprakel:2011,Grenard:2014,Stickland:2015} and oscillatory \cite{Gibaud:2010,Chan:2012,Perge:2014b} stress reveal that the fluidization process of attractive gels initially at rest, may take up to tens of hours. Experiments coupled to velocimetry have revealed that such a process is mostly activated, as evidenced by the Arrhenius-like dependence of the fluidization time with the applied shear stress \cite{Gibaud:2010,Grenard:2014}, and that it is strongly heterogeneous in both the vorticity and the flow directions \cite{Perge:2014b}. On the other hand, attractive gels show an overshoot in the stress response to shear startup experiments. Such behaviour, which strongly depends on the preshear history, corresponds to the orientation and subsequent rupture of the gel microstructure into clusters \cite{Varadan:2001,Mohraz:2005,Mock:2007,Rajaram:2010,Koumakis:2011}. Beyond the yield point, attractive gels either display homogeneous or shear-banded flows depending on the applied shear rate and on the boundary conditions \cite{Divoux:2016}. However, only a handful of studies have investigated the influence of the sample age, i.e. the duration separating preshear from the start of the experiment, on such shear-rate-induced fluidization scenario \cite{Raynaud:2002}. Recently, Martin and Hu have shown on Laponite suspensions \cite{Martin:2012} that aged-enough samples tend to exhibit long-lasting though transient shear-banding, and that shear-banding may also be trapped by the rapid aging of the non-flowing band and become permanent. The latter scenario is remarkable as it strongly differs from the classic shear-banding scheme which relies on the mechanical instability of the sample under scrutiny \cite{Ovarlez:2009,Fall:2010b}. This study highlights the interplay between the sample age and the shear and strongly urges to investigate the impact of sample age on the shear-induced fluidization scenario in other attractive gels.

To summarize, attractive gels neither entirely behave as solids nor as fluids over a wide range of timescales. In the landscape of non-Newtonian fluids, they define a rather singular category of materials rightfully referred to as the ``Twilight zone" in the early classification established by L.~Bilmes \cite{Bilmes:1942}, that was recently adapted to complex fluids by G.H.~McKinley \cite{McKinley:2015}. We barely start to understand the (transient) rheology of attractive gels and more experiments are obviously needed to shed some light on the heterogeneous flows of such highly time-dependent materials.

 In the present manuscript we report spatially resolved data on the fluidization dynamics of an attractive gel composed of weakly  attractive non-Brownian particles. Velocimetry performed in a concentric cylinder geometry simultaneously to shear startup experiments reveals that the steady-state behaviour is a subtle function of both the time $t_w$ during which the system was left to age before the beginning of the experiment, and of the value $\dot \gamma$ of the applied shear rate. Extensive experiments allow us to build a state diagram of steady-state flows in the ($\dot \gamma$, $t_w$) plane. Two distinct regions roughly emerge: ($i$)~homogeneous flows for shear rates larger than a critical value $\dot \gamma^*$ that weakly increases with the aging time, and ($ii$)~steady-state shear banding elsewhere. As a key result of this work, the complete fluidization observed in the upper region of the state diagram involves a transient shear band that is progressively eroded through a series of dramatic fluidization events. These avalanche-like processes show as large peaks in the temporal evolution of both the global shear stress and the slip velocity measured at the moving wall of the shear cell. As further confirmed by two-dimensional ultrasonic imaging, this fluidization process is spatially heterogeneous and may occur at different locations along the vorticity direction. Finally, for a range of parameters ($\dot \gamma$, $t_w$) in the vicinity of the boundary between the two main regions of the state diagram, we observe large fluctuations in the stress and slip velocity signals, although the system does not reach complete fluidization. Such avalanche-like events are strongly coupled to variations in both the width of the shear band and the slip velocity. The present work offers to our knowledge among the first experimental evidence of local avalanche-like fluidization events in a weak attractive gel under shear. It also provides an extensive data set to test the relevance of the flow stability criteria for shear banding \cite{Moorcroft:2013} and stands as a new challenge for spatially resolved models \cite{Colombo:2013,Colombo:2014,Fielding:2014}.

\section{Materials and Methods}

\subsection{Experimental setup}

\begin{figure*}[!t]
\centering
 \includegraphics[width=0.75\linewidth]{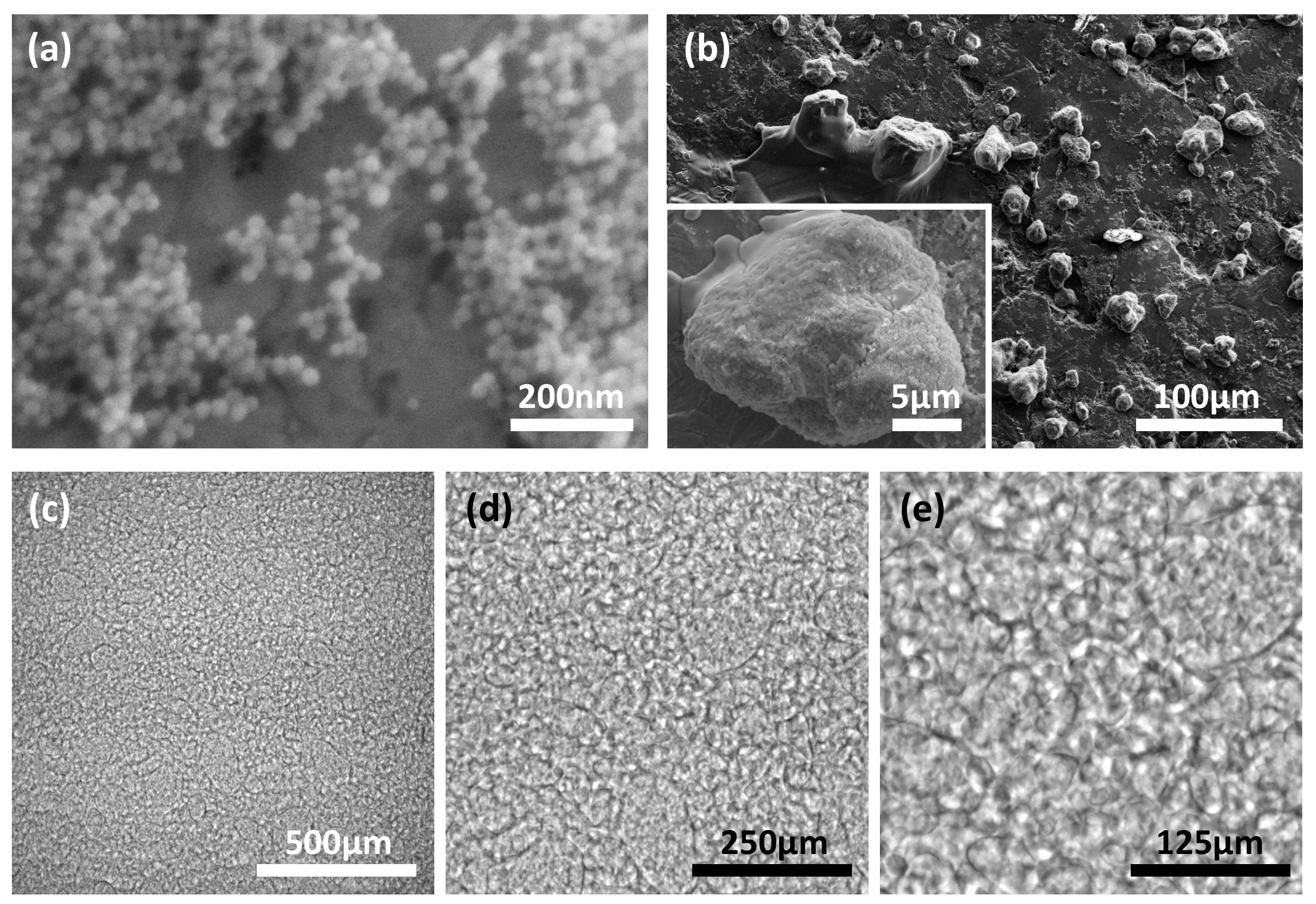}
  \caption{SEM images of (a) the stock colloidal suspension (Ludox TM-40) used to prepare the sample, and (b) the non-Brownian aggregates composing the sample which form almost instantaneously after mixing the colloidal suspension with a concentrated NaCl solution. The inset in (b) shows a single typical aggregate made of fused silica colloids. Images are obtained as follows: both the suspension and the gel have been diluted by a factor 1000 (except for the main figure in (b) where the gel has been diluted 100$\times$), the first one in deionized water and the second one in a 6.8\% wt. NaCl which corresponds to the final salt concentration in the gel. The samples are then left to dry on a pin stub before being imaged. (c)-(e) Bright-field microscopy images representative of the gel microstructure at three different magnifications. The sample has been presheared for 2~min at 500~s$^{-1}$ before being imaged about 10 minutes later in a homemade observation cell made of a glass slide topped with a cover slip and sealed with UV glue.}
  \label{fig1}
\end{figure*}

The experimental setup allows us to perform time-resolved velocimetry simultaneously to standard rheometry. The rheological measurements are performed in a polished Taylor-Couette (concentric cylinder) cell made of Plexiglas (gap $e=2$~mm) in which the inner Mooney-Couette cylinder of angle $2.3^{\circ}$, height 60~mm and radius 23~mm is connected to a stress-controlled rheometer (ARG2, TA Instruments) and positioned at $100~\mu$m from the bottom of the outer cylinder. A solvent trap located at the top of the rotor and a homemade lid are used to prevent evaporation efficiently up to about 9~hours. The Taylor-Couette cell is embedded in a water tank connected to a thermal bath which allows us to keep the sample at a constant temperature $T=23.5\pm 0.1^{\circ}$C.

Local velocity profiles across the gap of the Taylor-Couette cell are recorded simultaneously to the global rheology by means of two different ultrasonic probes immersed in the water tank, which also ensures acoustic coupling to the shear cell. The first ultrasonic probe is a single high-frequency focused transducer that allows us to record the azimuthal velocity $v$ as a function of the radial position $r$ across the gap at the middle height of the shear cell i.e. at 30~mm from the bottom. Full technical details on this one-dimensional ultrasonic velocimetry (1D-USV) technique can be found in a previous publication \cite{Manneville:2004a}. The second ultrasonic probe consists of a linear array of 128 transducers placed vertically at about 15~mm from the cell bottom. This transducer array is 32~mm high and gives access to images of the azimuthal velocity as a function of the radial position $r$ and vertical position $z$ over about 50\% of the cell height. This two-dimensional ultrasonic velocimetry (2D-USV) technique is thoroughly described in ref. \cite{Gallot:2013}. Both devices can be used simultaneously and roughly face each other in the water tank, i.e. they are separated by an azimuthal angle of about $180^{\circ}$. While the 1D-USV setup has the advantage of a better spatial resolution (about 40~$\mu$m against 100~$\mu$m), only the 2D-USV setup allows us to detect and monitor the presence of flow heterogeneities along the vorticity direction.

Both velocimetry techniques require that the ultrasonic beam crossing the gap of the shear cell is backscattered either by the fluid microstructure itself or by acoustic tracers added during sample preparation when the system is acoustically transparent \cite{Manneville:2004a,Gallot:2013}. Here, we shall emphasize that the microstructure of the sample further detailed below conveniently backscatters ultrasound in the single scattering regime, which allows us to monitor the fluid velocity in a fully non-invasive way.

\subsection{Sample preparation and rheological properties}
\label{Sampleprep}

\begin{figure}[t!]
\centering
 \includegraphics[width=\linewidth]{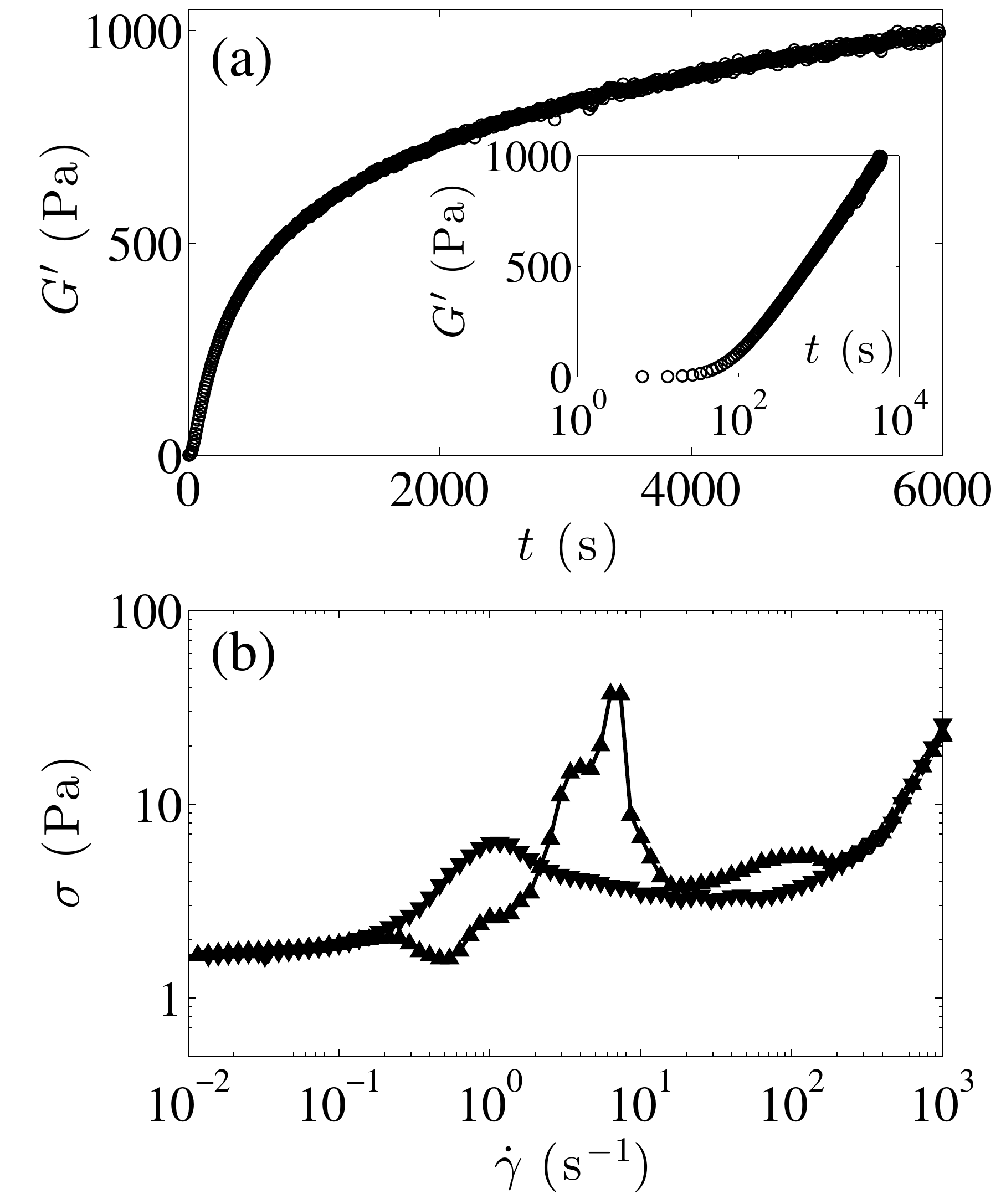}
  \caption{(a) Elastic modulus $G'$ vs time $t$ after a preshear at $\dot \gamma_p=500$~s$^{-1}$ for 2~min. Inset: same data set in semilogarithmic scales. Measurements performed under oscillatory shear stress in the linear regime with frequency 1~Hz and stress amplitude 0.05~Pa. (b)~Flow curve $\sigma$ vs $\dot \gamma$ obtained by first decreasing $\dot \gamma$ continuously from 10$^{3}$ to 10$^{-2}$~s$^{-1}$ in 75 logarithmically spaced points of duration $\delta t =8$~s ($\blacktriangledown$) and then increasing $\dot \gamma$ over the same range ($\blacktriangle$).}
  \label{fig2}
\end{figure}

Our Ludox gels are prepared following the same recipe as that used by M{\o}ller {\it et al.} in ref. \cite{Moller:2008}. As we shall show below, the system is composed of non-Brownian particles made of permanently fused silica colloids. These particles are themselves reversibly aggregated in brine due to van der Waals forces leading to a space-spanning gel with solid-like properties at rest.

A stable suspension of silica colloids (Ludox TM-40, Sigma-Aldrich, 40\% wt. in silica colloids) of typical diameter 20~nm [see Fig.~\ref{fig1}(a) for a Scanning Electron Microscopy (SEM) image of a dilute sample (Supra 55, VP Zeiss)] and of ${\rm pH}=9.0\pm0.5$ is first poured without any further purification into a 10\% wt. deionized aqueous solution of sodium chloride (Merck Millipore) up to a mass ratio Ludox:NaCl of 6:13, corresponding to a final volume fraction of 7\% in silica colloids and a final pH of $7.6\pm 0.2$. The mixture, which instantaneously becomes white and optically opaque, is then shaken intensely by hand for 2~min and left at rest for at least 15~h before being studied. Such a drastic change in the sample turbidity strongly suggests the rapid formation of aggregates at the micron-scale. Indeed, direct observations using different techniques confirm the existence of a much coarser microstructure than the initial nanometric silica colloids. On the one hand, SEM images of a dried droplet extracted from a fresh sample that has been previously diluted in a NaCl solution unveil the presence of particles which size ranges from a few microns up to a hundred microns [Fig.~\ref{fig1}(b)]. On the other hand, bright-field microscopy images (ECLIPSE Ti, Nikon) of the sample neither altered nor diluted further confirm the existence of these micron-sized particles [Fig.~\ref{fig1}(c)-(e)], which are stable in time and robust to repeated shear, as confirmed by similar observations performed on samples submitted to different shear histories (images not shown). 

To account for the formation of such a large scale microstructure that to our knowledge has not been reported in the literature previously \cite{Kobayashi:2005,Moller:2008}, we propose the following scenario. Above pH=7, silica colloids are negatively charged and bear silanol (Si$-$OH) and dissociated silanol groups that are poorly hydrated \cite{Heston:1960}. In most previous works, NaCl is added in a relatively small amount (typically 0.05--0.5~M) such that electrostatic repulsion is screened leading to the slow reversible aggregation of individual colloids up to the formation of a colloidal gel \cite{Trompette:2004,Cao:2010,Truzzolillo:2014pp}. Here, we add a much larger amount of salt (1.2~M) to the colloidal suspension, which leads to an ion exchange where protons are replaced by sodium ions \cite{Allen:1969,Allen:1970}. The loss of hydrogen bonding between the colloids and the solvent triggers the fast and irreversible aggregation of the silica colloids through the formation of interparticle siloxane bonds \cite{Depasse:1970,Depasse:1997,Drabarek:2002}, resulting in the formation of the non-Brownian particles described above. Finally, these non-Brownian particles also aggregate reversibly due to van der Waals forces and form a space-spanning network, i.e. a gel, which mechanical behaviour is studied below. Note that such microstructure scatters ultrasound efficiently, allowing us to use both 1D- and 2D-USV without requiring any seeding of the sample with tracer particles. 

The rheological features of the gel are displayed in Fig.~\ref{fig2}. A strong preshear of $\dot \gamma_p=500$~s$^{-1}$ applied during 2~min fully fluidizes the system, which quickly rebuilds once the preshear is stopped and subsequently shows pronounced aging. Indeed, small amplitude oscillatory shear reveals that the elastic modulus $G'$ becomes larger than the viscous modulus $G''$ within about 20~s [see Fig.~1 in the ESI] and that $G'$ further increases logarithmically over more than 2~h [Fig.~\ref{fig2}(a)]. Such aging dynamics are reproducible for a given preshear protocol and lead to the formation of a solid-like gel. The latter shows an elastic modulus that is nearly frequency independent and a critical yield strain of about 7\% that weakly depends on the sample age [see Figs.~2 and 3 in the ESI].

Such a solid-like behaviour also reflects in the nonlinear rheology of the gel. Fig.~\ref{fig2}(b) shows the flow curve $\sigma$ vs $\dot \gamma$ measured by sweeping down the shear rate from $10^{3}$ to $10^{-2}$~s$^{-1}$ and back up. The flow curve shows an apparent yield stress of a few Pascals and displays a complex non-monotonic shape together with strong hysteresis. Velocity profiles recorded simultaneously to the up and down shear rate sweeps shown in Fig.~\ref{fig2} reveal that wall slip and heterogeneous flows are involved over large ranges of shear rates, below 20~s$^{-1}$ during the decreasing ramp and up to about 200~s$^{-1}$ during the increasing ramp [see Fig.~4 in the ESI]. In particular, the large stress peak observed in Fig.~\ref{fig2}(b) at $\dot \gamma \simeq 7$~s$^{-1}$ while ramping up the shear rate is the signature of the partial fluidization of the sample which moves as a plug and totally slips at both walls for $7\lesssim \dot \gamma \lesssim 100$~s$^{-1}$, before finally starting to flow homogeneously on the reversible branch at high shear rates i.e. above about 200~s$^{-1}$.

\begin{figure}[t!]
\centering
 \includegraphics[width=\linewidth]{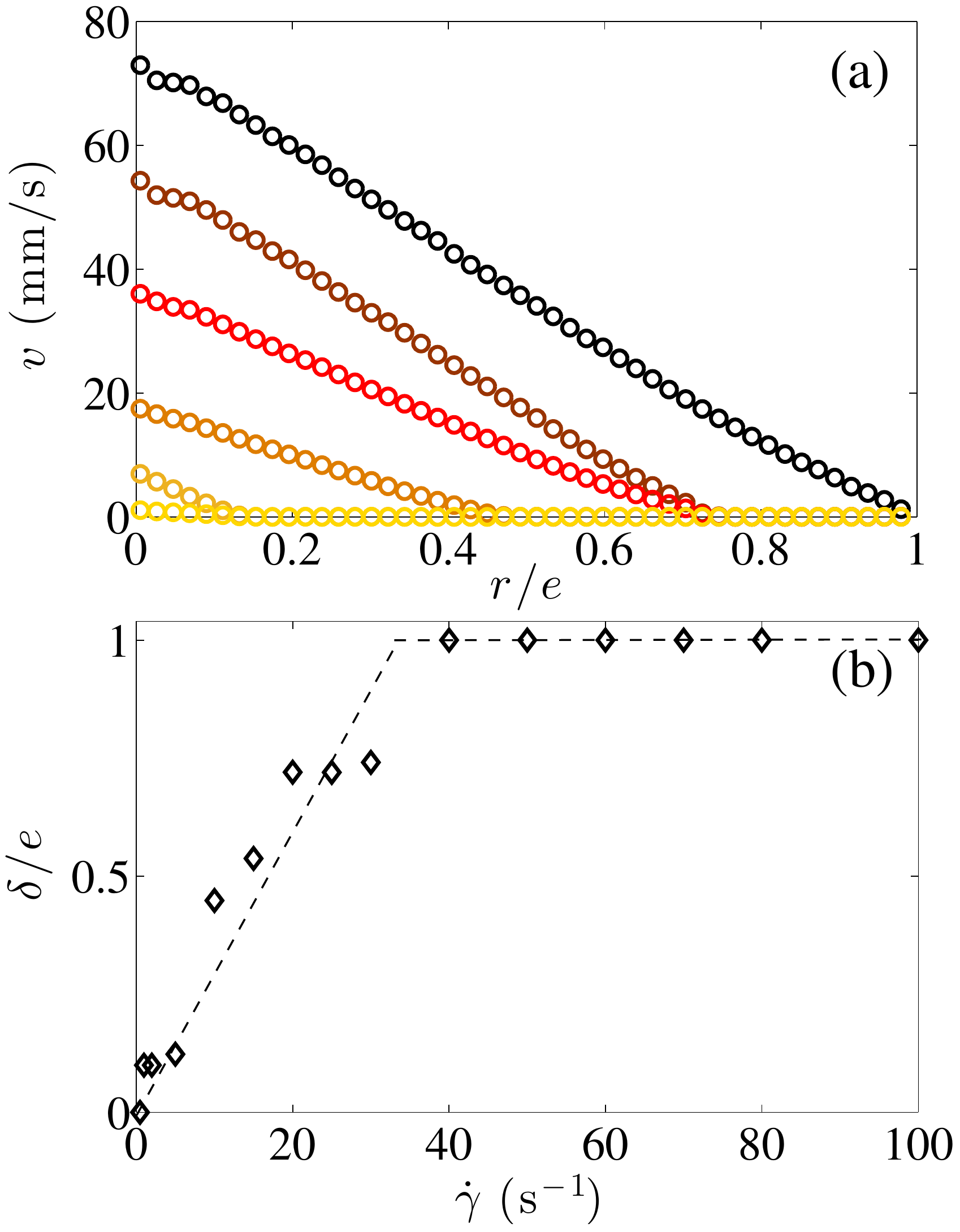}
  \caption{(a) Steady-state velocity profiles across the gap at different shear rates: $\dot \gamma = 40$, 30, 20, 10, 5 and 1~s$^{-1}$ from top to bottom. The parameter $r$ denotes the distance to the rotor. These data have been recorded starting from large shear rates (500~s$^{-1}$) and decreasing $\dot \gamma$ by successive steps of long enough duration to achieve a steady state at every imposed shear rate. (b) Size of the fluidized band $\delta$ normalized by the gap width $e$ vs the applied shear rate $\dot \gamma$. The dashed line corresponds to the standard ``lever rule'' with $\dot \gamma_c=35$~s$^{-1}$.}
  \label{fig2b}
\end{figure}

Furthermore, in Fig.~\ref{fig2}(b), the shear stress $\sigma$ shows a minimum in between $\dot \gamma \simeq 20$~s$^{-1}$ and 40~s$^{-1}$ on the decreasing shear rate sweep, which hints at the existence of a critical shear-rate $\dot \gamma_c$ below which no homogeneous flow is possible \cite{Coussot:2002a,Ragouilliaux:2006}. The latter result is confirmed by performing steady-state measurements at constant applied shear rate starting from the liquid state, i.e. on a fully fluidized sample, in order to avoid long-lasting transients that go along with shear startup experiments on a sample at rest, and which are at the core of section~\ref{results}. Here, discrete shear rates of decreasing values ranging from 500~s$^{-1}$ down to 0.1~s$^{-1}$ are successively applied for at least 300~s each. The flow remains homogeneous down to $\dot \gamma_c \sim 35$~s$^{-1}$ below which the sample exhibits an arrested band [Fig.~\ref{fig2b}(a)]. Such a value of $\dot \gamma_c$ is comparable to the one reported in a previous work on very similar Ludox gels \cite{Moller:2008}. Moreover, as the shear rate is decreased below $\dot \gamma_c$, the size of the fluidized band $\delta$ decreases roughly linearly with $\dot \gamma$ [Fig.~\ref{fig2b}(b)] in agreement with the classical ``lever rule'' \cite{Coussot:2002a,Ragouilliaux:2006,Moller:2008}. The deviation of $\delta$ from a straight line could be related to the wall slip that is present at the rotor. 

To conclude this subsection, complex cycles of hysteresis as the one reported in Fig.~\ref{fig2} have been also reported for numerous other attractive gels including carbon black gels \cite{Grenard:2014} and clay suspensions \cite{tenBrinke:2007,Gibaud:2009}. Although significant progress has been made in extracting quantitative information from hysteresis loops \cite{Divoux:2013}, we rather chose to focus our study on shear startup experiments in order to fully decouple the fluid dynamics from any time-dependent effect related to the experimental protocol. The present experiments are thus all performed on a sample prepared in a solid-like state, so as to investigate the shear-induced fluidization scenario of the non-Brownian particle gel. In practice, after preparing the gel in a well-defined and reproducible initial state thanks to preshearing, we monitor the stress response of the material to a constant shear rate over long durations. The results are discussed in section~\ref{results}.

\begin{figure*}[!t]
\centering
  \includegraphics[width=0.7\linewidth]{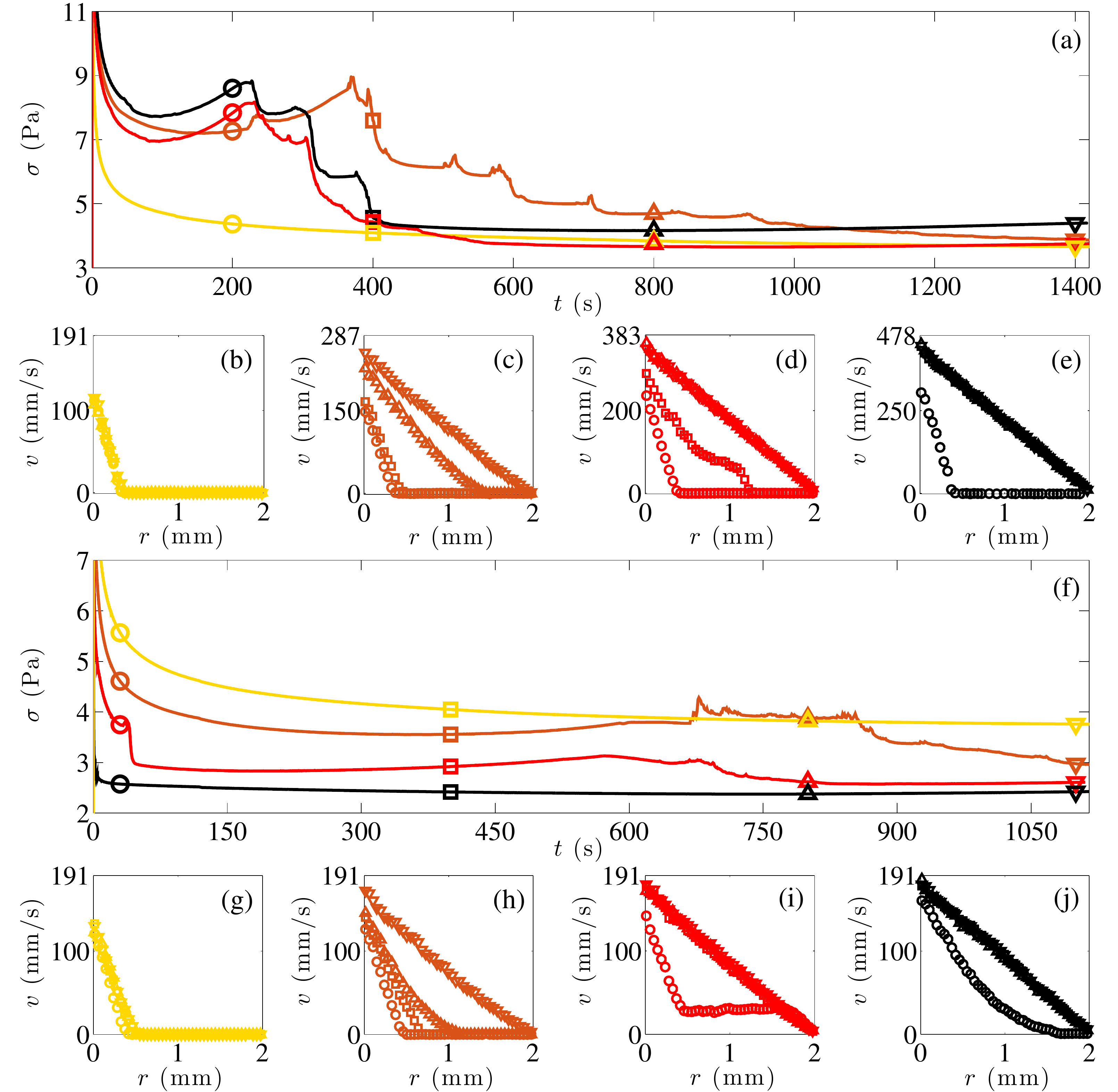}
  \caption{
 (a) Shear stress response $\sigma (t)$ for different applied shear rates [color, $\dot \gamma$ (s$^{-1}$)]: [\textcolor{yellow}{\textbf{\large --}}, 100]; [\textcolor{orange}{\textbf{\large --}}, 150]; [\textcolor{red}{\textbf{\large --}}, 200]; [\textcolor{black}{\textbf{\large --}}, 250]. (b)-(e) Velocity profile $v(r)$ across the gap, where $r$ is the distance to the rotor at different times [symbol, time (s)]: [$\bigcirc$, 200]; [$\Box$, 400]; [$\bigtriangleup$, 800]; [$\bigtriangledown$, 1400]. Each color corresponds to each applied shear rate in (a). The rotor velocity corresponds to the upper bound of the vertical axis. The sample is left to age during $t_w = 60$~min before each experiment. 
 (f) Shear stress response $\sigma (t)$ for different waiting times $t_w$ [color, $t_w$ (min)]: [\textcolor{yellow}{\textbf{\large --}}, 100]; [\textcolor{orange}{\textbf{\large --}}, 30]; [\textcolor{red}{\textbf{\large --}}, 5]; [\textcolor{black}{\textbf{\large --}}, 1]. Experiments  performed at $\dot\gamma=$~100~s$^{-1}$. (g)-(j) Velocity profile at different times [symbol, time (s)]: [$\bigcirc$, 30]; [$\Box$, 400]; [$\bigtriangleup$, 800]; [$\bigtriangledown$, 1100]. Each color corresponds to each value of $t_w$ in (f). The rotor velocity corresponds to the upper bound of the vertical axis.
}
    \label{fig3}
\end{figure*}

\subsection{Experimental protocol}

Prior to any shear startup experiment, the sample is presheared at $\dot \gamma_p=500$~s$^{-1}$ for 2~min in order to erase any previous shear history \cite{Viasnoff:2002}. We check systematically that the velocity profiles during the preshear step become homogeneous across the gap within a duration shorter than 2~min. The system is then left to rebuild for a duration $t_w$ that ranges between 1 and 100~min, and during which we apply small amplitude oscillatory shear stress (stress amplitude $\sigma=0.05$~Pa, frequency $f=1$~Hz) to monitor the evolution of the gel linear viscoelastic properties and make sure that for a given value of $t_w$, the gel reaches an initial state that is reproducible from one experiment to another. Finally, a constant shear rate $\dot \gamma$ is applied to the material over a long duration ($>1000$~s) while we monitor the stress response together with local velocity profiles.

\section{Results}
\label{results}

\subsection{Global rheology and one-dimensional velocity profiles}

We first fix the duration between the preshear and the start of the experiment to $t_w=60$~min and discuss the material response for shear startup experiments performed at different shear rates. For shear rates lower than about $150$~s$^{-1}$, the stress relaxes smoothly towards a constant value as illustrated in figure~\ref{fig3}(a) for an experiment performed at $\dot \gamma=100$~s$^{-1}$. Velocity profiles acquired simultaneously reveal that the material remains at rest near the stator, while it flows in the vicinity of the rotor where it shows strong wall slip [Fig.~\ref{fig3}(b)]: the sample displays steady-state shear banding. For larger shear rates, one observes a completely different scenario, both from global and local measurements. The stress displays a series of spike-like relaxation events, during which stress increases of small amplitude are followed by large drops [Fig.~\ref{fig3}(a)]. Local measurements show that the gel flows heterogeneously. However, the shear band is only transient and the system always ends up being homogeneously sheared at steady state [Fig.~\ref{fig3}(c)--(e)]. Qualitatively, the width of the shear band increases while strong wall slip is observed at the rotor, until the whole gap is fluidized and wall slip becomes negligible. This scenario is robustly observed at different shear rates and full fluidization occurs sooner for larger shear rates [Fig.~\ref{fig3}(a)]. A quantitative analysis of the velocity profiles is further proposed in section~\ref{localanalysis}. 

\begin{figure}[t!]
\centering
  \includegraphics[width=\linewidth]{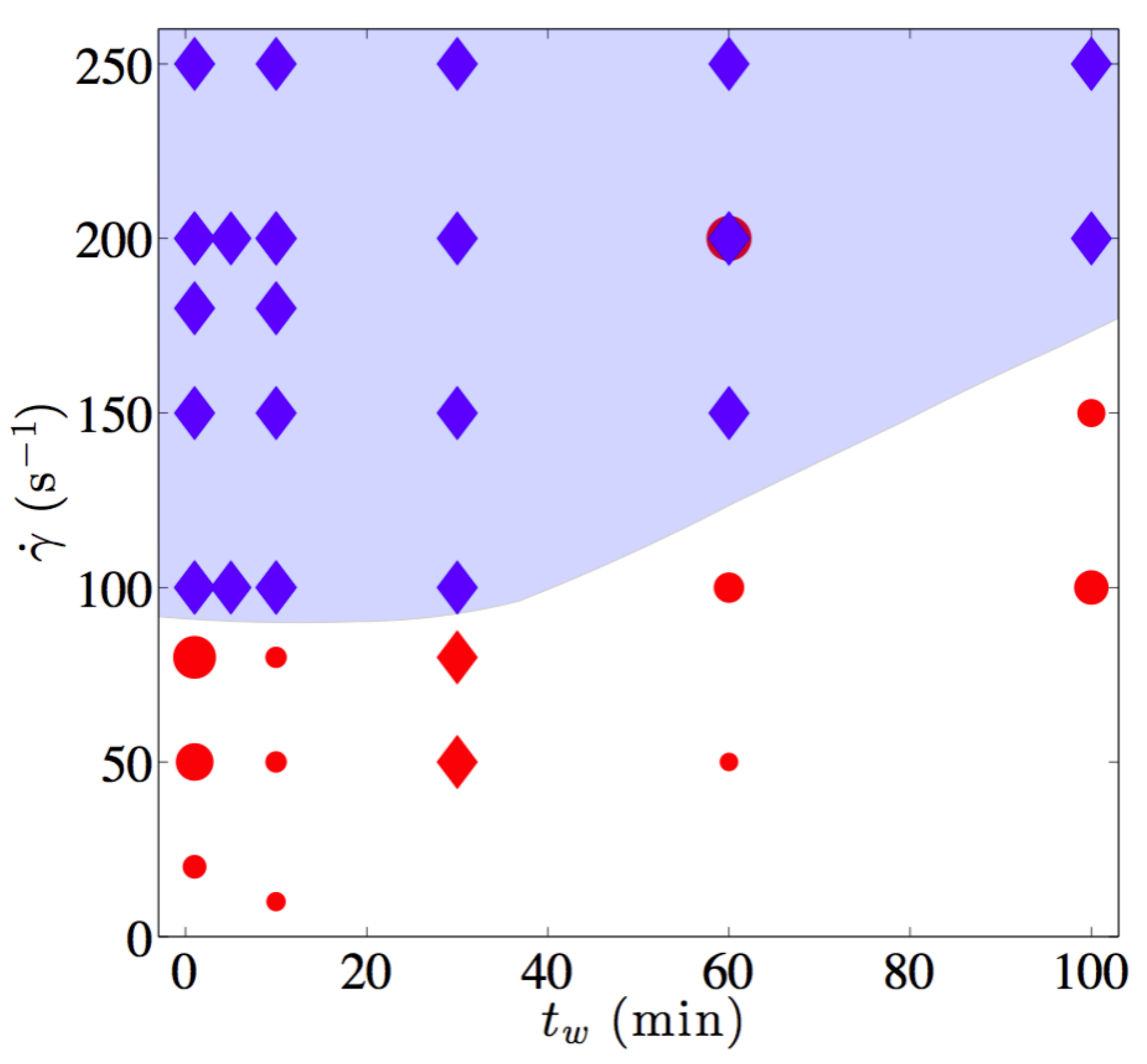}
  \caption{Flow state diagram in the (sample age $t_w$, applied shear rate $\dot\gamma$) plane. In steady state, the gel may either be fully fluidized (\textcolor{blue}{$\blacklozenge$}, blue region) or display shear banding. Steady shear banding is represented in (\textcolor{red}{$\bullet$}) symbol which size encodes the portion of the gap that is being sheared. Unsteady banding, which denotes flows where the band width and the slip velocity at the rotor display significant fluctuations, is represented by (\textcolor{red}{$\blacklozenge$}).
  }
  \label{fig4}
\end{figure}

To evidence the impact of the aging time $t_w$ on the material response, Figs.~\ref{fig3}(f)--(j) show similar experiments performed for different $t_w$ at the same shear rate $\dot \gamma=100$~s$^{-1}$. For long waiting times between the preshear and the start of the experiments, e.g. $t_w=100$~min, one observes a smooth stress relaxation here again associated with steady shear-banded velocity profiles [Fig.~\ref{fig3}(f) and (g)]. Decreasing $t_w$ to 30 or 5~minutes leads to a gel of lower elastic modulus. Local measurements further reveal that these weaker gels go through a transient shear-banding regime and that in both cases the steady state is homogeneous [Fig.~\ref{fig3}(h) and (i)]. Here, unlike the case of transient shear banding reported in Fig.~\ref{fig3}(a) for large shear rates in samples left to age for $t_w=60$~min where fluidization corresponds to a long series of successive stress relaxations, fluidization proceeds in a single stress drop together with small noisy fluctuations [Fig.~\ref{fig3}(f)]. Finally, one observes that very young gels ($t_w=1$~min) barely show any heterogeneous velocity profile during startup flow and reach a homogeneous steady state within a few tens of seconds [Fig.~\ref{fig3}(j)]. In summary, the longer the sample ages after preshear, the more likely it is to exhibit a long-lasting heterogeneous fluidization scenario or to display steady shear banding.

\begin{figure}[t!]
\centering
  \includegraphics[width=\linewidth]{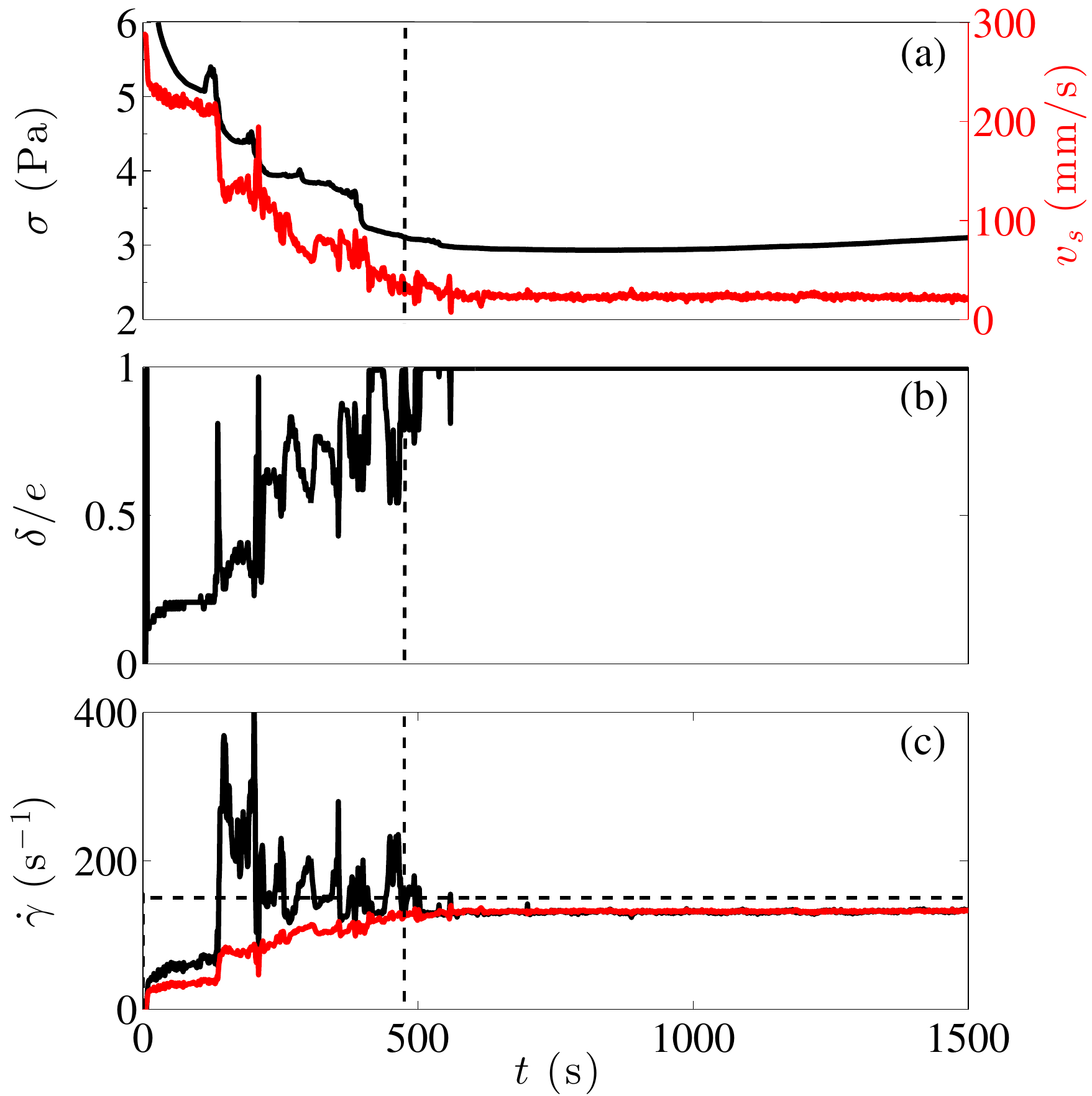}
  \caption{Full fluidization observed for $\dot\gamma=150$~s$^{-1}$ and $t_w=30$~min. (a) Shear stress $\sigma$ (\textcolor{black}{\textbf{\large --}}) and slip velocity $v_s$ (\textcolor{red}{\textbf{\large --}}~) vs time. (b)~Width $\delta$ of the fluidized shear band normalized by the gap width $e$
vs time. (c)~Local shear rate within the shear band (\textcolor{black}{\textbf{\large --}}) and global shear rate (\textcolor{red}{\textbf{\large --}}) \break vs time. The horizontal dotted line indicates the shear rate applied by the rheometer. In (a)-(c), the vertical dashed line indicates the fluidization time $\tau_f$.}
  \label{fig5}
\end{figure}

\begin{figure}[!t]
\centering
  \includegraphics[width=\linewidth]{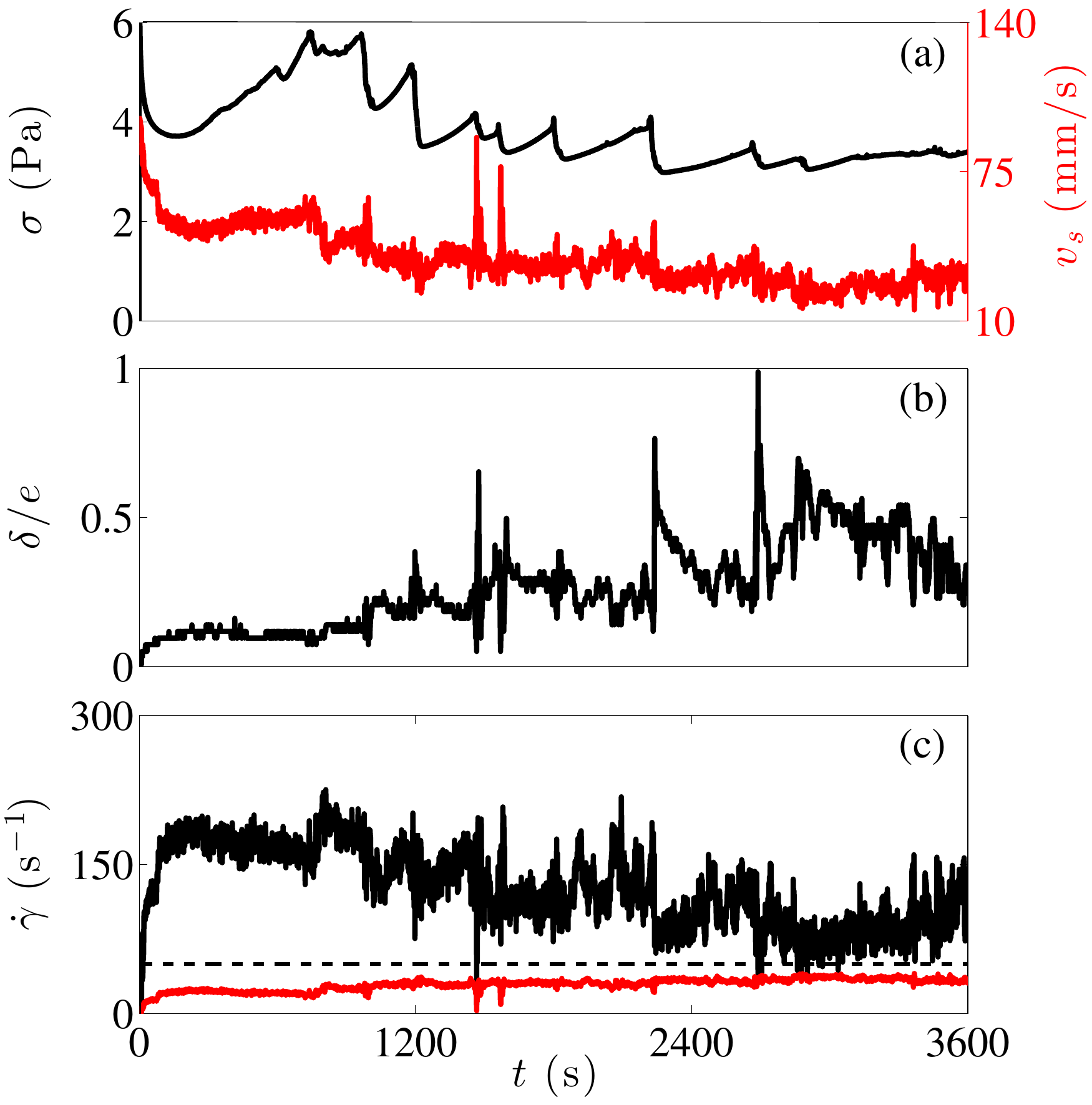}
  \caption{Unsteady shear banding observed for $\dot\gamma=50$~s$^{-1}$ and $t_w=30$~min. (a) Shear stress $\sigma$ (\textcolor{black}{\textbf{\large --}}) and slip velocity $v_s$ (\textcolor{red}{\textbf{\large --}}) vs time. (b)~Width $\delta$ of the fluidized shear band normalized by the gap width $e$
vs time. (c)~Local shear rate within the shear band (\textcolor{black}{\textbf{\large --}}) and global shear rate (\textcolor{red}{\textbf{\large --}}) vs time. The horizontal dotted line indicates the shear rate applied by the rheometer.}
  \label{fig6}
\end{figure}

\subsection{Flow state diagram}
\label{PD}

To go one step further and get an overall picture of the material response, we have performed extensive shear startup experiments by varying systematically both the aging time ($1\leq t_w \leq 100$~min) and the applied shear rate ($10 \leq \dot \gamma \leq 250$~s$^{-1}$). The entire data set is summarized in the flow state diagram pictured in Fig.~\ref{fig4}, where the different symbols code for the three different scenarios that we may distinguish at the end of each startup experiment lasting typically 1500 to 3600~s. In steady state, the attractive gel may either fully fluidize after a transient phase possibly involving strong fluctuations (\textcolor{blue}{$\blacklozenge$} in Fig.~\ref{fig4}), as observed in the upper part of the state diagram for large shear rates quite independently of the sample age, or exhibit shear banding. 
The boundary between these two regimes defines a critical shear rate $\dot \gamma^*$ that depends in a nonlinear fashion on the sample age $t_w$. Furthermore, we can discriminate between two types of behaviour when shear banding is observed.

First, \textit{steady} shear banding (\textcolor{red}{$\bullet$}~in Fig.~\ref{fig4}), for which both the width $\delta$ of the shear band and the slip velocity $v_s$ at the rotor display negligible fluctuations in steady state, is observed in particular at low aging times $t_w \lesssim 20$~min and under low enough shear rates, typically below 100~s$^{-1}$. In that case, the width of the arrested band may decrease ($t_w=1$~s) or remain constant ($t_w=10$~s and 100~s) for increasing shear rates.

Second, we have also observed \textit{unsteady} shear banding (\textcolor{red}{$\blacklozenge$}~in Fig.~\ref{fig4}). In this case both global and local measurements display significant fluctuations in steady state. These fluctuations are strikingly similar to those observed during the transients leading to complete fluidization in the upper part of the diagram. However here the material never gets entirely fluidized and  the shear band width $\delta$ does not show a systematic evolution towards $\delta=e$ so that an unsteady shear band persists in steady state at least within the finite duration of the experiments. These fluctuations are investigated in more details in the next section.

\subsection{Linking global rheology to the local dynamics}
\label{localanalysis}

 In this section, we focus on the strong fluctuations that are observed (i)~during transient regimes leading to complete fluidization and (ii)~during unsteady shear-banded flows at steady state. Global rheological data together with the detailed analysis of the corresponding local measurements are reported in Figs.~\ref{fig5} and \ref{fig6} respectively (see also corresponding Movies~1 and 2 in the ESI). Local data are analyzed as follows: linear fits of the velocity profiles in the fluidized part of the gap are used to estimate the local shear rate and to extrapolate the sample velocity $v(0)$ at the rotor. The width $\delta$ of the shear band is obtained as the abscissa of the intersection between the fit and the zero velocity axis, while the slip velocity is given by $v_s=v_0-v(0)$, where $v_0$ is the rotor velocity. Finally, the global shear rate is defined as $v(0)/e$.
  
We first discuss Fig.~\ref{fig5} which shows a case of full fluidization for $\dot \gamma=150$~s$^{-1}$ on a gel left to age for $t_w=30$~min. The stress relaxes by successive jumps up to full fluidization which occurs at $t=\tau_f \simeq 507$~s and after which $\sigma$ remains roughly constant [Fig.~\ref{fig5}(a)]. The temporal evolutions of $\delta$ [Fig.~\ref{fig5}(b)] and of the local shear rate [Fig.~\ref{fig5}(c)] show that fluidization occurs by successive spatial ``avalanches'' that are directly correlated to the stress drops. During the whole fluidization process, the slip velocity at the rotor $v_s$ decreases towards a negligible value that is reached at  $t=\tau_f$, diminishing by jumps that are in phase with the stress evolution.

Figure~\ref{fig6} focuses on spatiotemporal fluctuations observed during steady-state shear-banded flows for the same aging time ($t_w=30$~min) as in Fig.~\ref{fig5} but sheared at a lower shear rate ($\dot \gamma=50$~s$^{-1}$). The stress exhibits periods of slow increase followed by rapid drops [Fig.~\ref{fig6}(a)]. Within an hour, about half the gap gets fluidized and the size of the fluidized band shows pronounced increase during short period of times, that are synchronized with the stress drops [Fig.~\ref{fig6}(b)]. The dynamics of the fluid at the rotor is strongly correlated to the global fluctuations, as evidenced by the sudden peaks of the slip velocity when the stress drops. Note that, although the slip velocity decreases in average over the whole duration of the experiment, it remains at a high level of about 20\%, in noticeable contrast with the fully fluidized scenario described in the previous paragraph. Such oscillatory dynamics in the vicinity of the rotor are reminiscent of stick-slip. Indeed, the fluidized band shows time interval during which the local shear rate remains constant [Fig.~\ref{fig6}(c)], while the band width slowly decreases. As a result, the stress slowly builds up, until the fluidized band experiences a large slip event at the moving wall and gets rejuvenated. These dynamics look similar to the stick-slip motion reported in clays in the pioneering work of Pignon {\it et al.} \cite{Pignon:1996}. However, in the latter case, stick-slip occurs along fracture planes located in the bulk sample, while here slip at the wall appears to play a key role.

Finally, we shall emphasize that in both types of scenarios, the peaks in both the stress and the slip velocity are only seen in the presence of shear banding. Although it is not clear which of the shear banding or of the slip at the rotor is the cause of the oscillations, these peculiar dynamics hint toward a subtle flow--microstructure coupling that certainly deserves more investigation.

\begin{figure}[t!]
\centering
  \includegraphics[width=\linewidth]{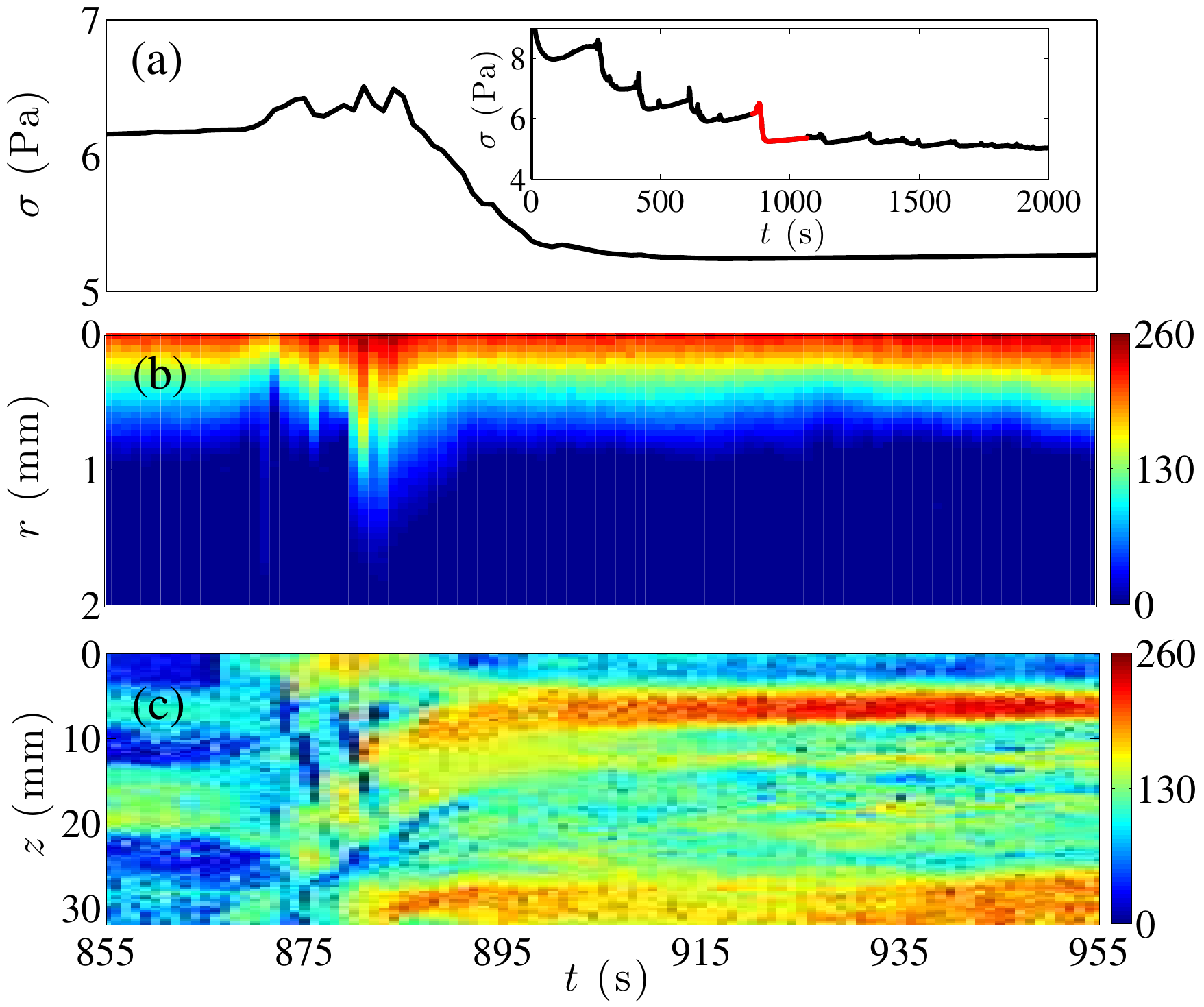}
 \caption{(a) Stress $\sigma$ vs time during a single stress drop event extracted from shear startup experiment during which the material is fully fluidized ($\dot\gamma=150$~s$^{-1}$, $t_w=10$~min). Inset: stress vs time for the whole experiment. The signal pictured in the main graph appears in red. (b)~Spatiotemporal diagram of the velocity data $v(r,t)$ as a function of position $r$ and time $t$. The radial position $r$ is measured from the rotating inner wall. Data obtained with 1D-USV. (c)~Spatiotemporal diagram of the velocity data $v(r_0,z,t)$ as a function of the vertical position $z$ and time $t$ at $r_0=0.2$~mm. Data obtained with 2D-USV. The vertical position $z$ is measured from the top of the transducer array. In both (b) and (c), the fluid velocity is color coded in mm.s$^{-1}$.}
  \label{fig7}
\end{figure}

\subsection{Local scenario within an avalanche}

\begin{figure}[t!]
\centering
  \includegraphics[width=\linewidth]{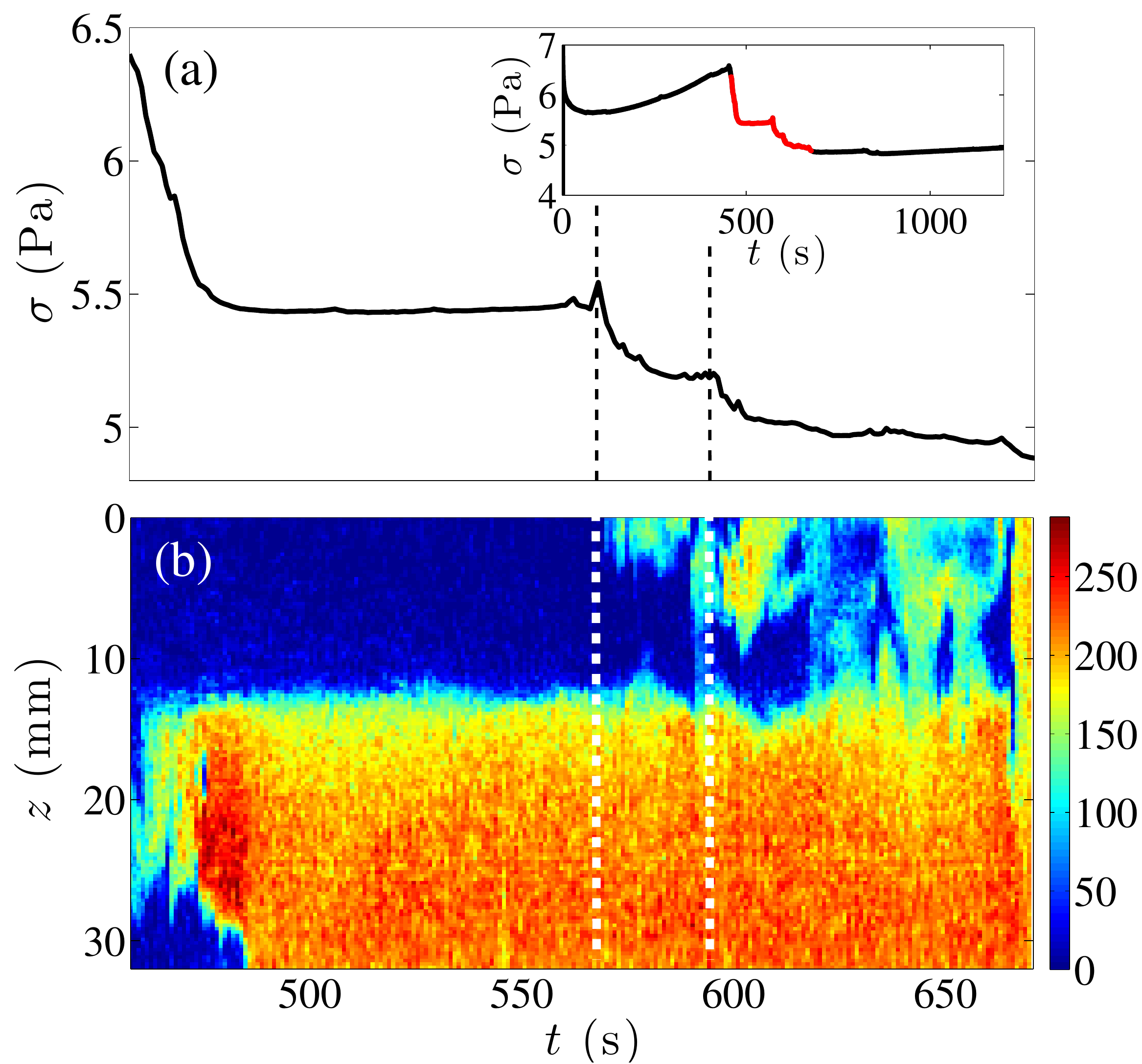}
   \caption{(a) Stress $\sigma$ vs time during a single stress relaxation episode extracted from shear startup experiment at the end of which the material is fully fluidized ($\dot\gamma=180$~s$^{-1}$, $t_w=1$~min). (b) Spatiotemporal diagram of the velocity data $v(r_0,z,t)$ as a function of the vertical position $z$ and time $t$ at $r_0=0.7$~mm. Data obtained with 2D-USV. The vertical position $z$ is measured from the top of the transducer array. The fluid velocity is color coded in mm.s$^{-1}$.}
  \label{fig8}
\end{figure}

\begin{figure}[t!]
\centering
  \includegraphics[width=8.6cm]{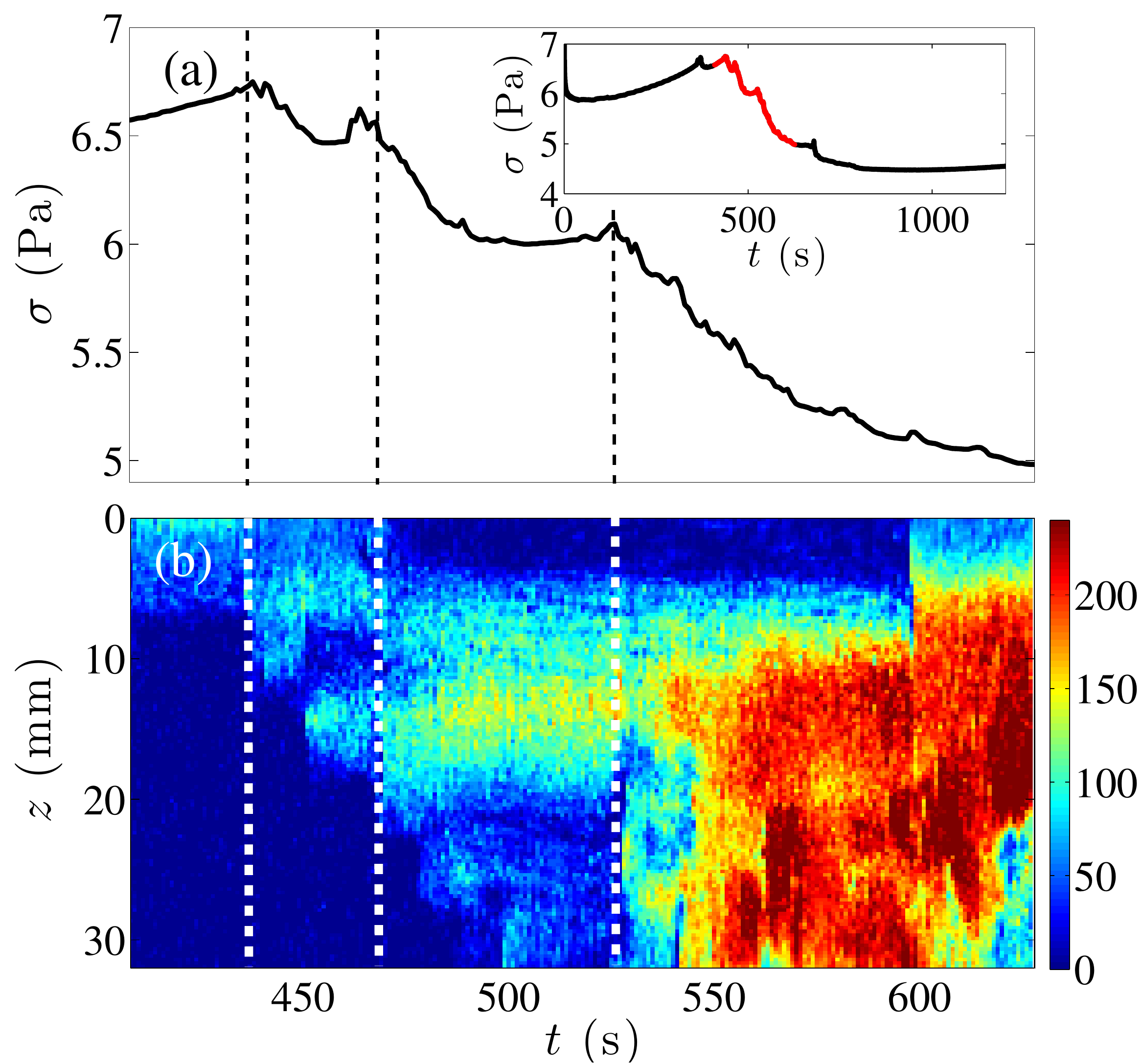}
  \caption{(a) Stress $\sigma$ vs time during a single stress relaxation episode extracted from shear startup experiment at the end of which the material is fully fluidized ($\dot\gamma=200$~s$^{-1}$, $t_w=1$~min). (b) Spatiotemporal diagram of the velocity data $v(r_0,z,t)$ as a function of the vertical position $z$ and time $t$, at $r_0=0.7$~mm. Data obtained with 2D-USV. The vertical position $z$ is measured from the top of the transducer array. The fluid velocity is color coded in mm.s$^{-1}$.}
  \label{fig9}
\end{figure}

To get better insight on the avalanche-like fluidization scenario, we now focus on the single stress drop shown in Fig.~\ref{fig7}(a). This event is extracted from a shear startup experiments performed at $\dot \gamma =150$~s$^{-1}$ for $t_w=10$~min and the steady-state flow corresponds to a fully fluidized sample [inset of Fig.~\ref{fig7}(a)]. The velocity profiles recorded simultaneously with 1D-USV are plotted as a spatiotemporal diagram in Fig.~\ref{fig7}(b). Quite surprisingly the one-dimensional flow profiles before and after the stress drop are very similar as  they all show shear localization over about half the gap. By the time the stress reaches its maximum value, about 80\% of the gap is sheared. Then shear abruptly localizes again over about 1~mm close to the rotor at the beginning of the stress relaxation. Therefore, although the strong fluctuations of $v(r,t)$ observed before the stress peak appear to be correlated to those of $\sigma(t)$, the drop of about 15\% in the stress value cannot be explained by these one-dimensional data in a straightforward manner. We emphasize that the 1D-USV measurements are performed at a given height of the Taylor-Couette cell so that they may reflect the evolution of global rheology only if the flow is homogeneous along the vorticity direction. As a matter of fact, velocity profiles $v(r,z,t)$ recorded simultaneously over the whole height of the Taylor-Couette cell through 2D-USV demonstrate that the flow is strongly heterogeneous in the vertical direction $z$. Figure~\ref{fig7}(c) shows a spatiotemporal plot of the fluid velocity along the vertical axis $z$ at a fixed radial position $r_0=0.2$~mm close to the rotor (see also Movie~3 in the ESI). Depending on the position along the $z$-axis, the material can simultaneously be either solid-like as evidenced by areas of very low velocities in Fig.~\ref{fig7}(c) (see for instance at the top of the region of interest for $z\lesssim 5$~mm) or fluid-like and flow at high velocities as observed for $z\simeq 8$ and $z\simeq 30$~mm. Note that the 1D-USV measurements are performed at $z_0\simeq 15$~mm and are fully consistent with the 2D-USV data. Our measurements further reveal that the particular avalanche-like event analyzed here only corresponds to partial fluidization: at the end of the stress drop, the flow is still heterogeneous, even though arrested regions for $t\lesssim 870$~s has given way to flowing regions arranged in a vertically banded structure for $t\gtrsim 890$~s. In fact, the inset of Fig.~\ref{fig7}(a) shows that several subsequent stress drop events still have to take place before complete fluidization of the sample.

To test the existence of a local fluidization scenario that would be generic to all avalanche-like events, we analyze additional experiments in the full fluidization region of the state diagram at $\dot \gamma =180$ and 200~s$^{-1}$. The temporal evolution of the shear stress is reported respectively in Figs.~\ref{fig8}(a) and \ref{fig9}(a) and the corresponding 2D-USV measurements $v(r_0,z,t)$ are displayed as spatiotemporal diagrams in Figs.~\ref{fig8}(b) and \ref{fig9}(b) for a radial position $r_0=0.7$~mm (see also Movies~4 and 5 in the ESI). Here again, the temporal evolution of the velocity field is strongly heterogeneous along the vertical direction in both experiments. Furthermore, despite these two experiments are performed at comparable shear rates, the local fluidization scenario is strikingly different. In the shear startup experiment reported in Fig.~\ref{fig8}, the lower part of the region of interest is fully fluidized after the first stress relaxation while the upper part of the sample necessitates two supplemental avalanches to in turn fully fluidize. On the other hand, the fluidization process shown in Fig.~\ref{fig9} starts from the upper part of the region of interest before extending to the lower part of the Taylor-Couette cell. Here, each avalanche-like event involves large pieces of the sample with a typical vertical extension of about 5~mm [see events marked by white dotted lines in Fig.~\ref{fig9}(a)]. Some of these events only show up in the stress response as very small peaks while their local signature is much more impressive (see, e.g., the event at $t\simeq 600$~s). For $t\simeq 400$ to 500~s, the sample even appears to fluidize, or at least to be set into motion, through regular steps occuring from top to bottom with a characteristic time of 10--20~s. The experiments shown in Figs.~\ref{fig8} and \ref{fig9} allow us to conclude that the peaks in the stress signal may encompass very different local scenarios. Since both experiments were performed with the same aging time $t_w=1$~min, these results illustrate the high sensitivity to ``initial conditions,'' i.e. to the possibly different arrangement of the heterogeneous microstructure after the aging process, and the subtle interplay between aging and shear banding with no systematic failure scenario along the cell height.

\section{Discussion}

On the one hand, we conclude from section~\ref{Sampleprep} that the gel under scrutiny displays a discontinuous yielding transition that is very similar to the one reported in other attractive gels \cite{Coussot:2002c,Ragouilliaux:2006,Moller:2008}. Indeed, one observes that, when the initial condition is a completely fluidized state, applying a series of shear rates of decreasing values leads to the growth of a steady-state arrested band below a critical shear rate $\dot \gamma_c$. The value of $\dot \gamma_c$ is inherent to the sample and results from the flow instability at low shear rates. This is fully consistent with previous results on Ludox gels. \cite{Moller:2008} On the other hand, the shear startup experiments reported in section~\ref{results} and performed on the sample prepared in the solid state allow us to identify a second critical shear rate $\dot \gamma^*>\dot \gamma_c$. For applied shear rates such as $\dot \gamma _c< \dot \gamma<\dot \gamma^*$, the sample gets only partially fluidized and exhibits steady-state shear banding, while long-lasting yet transient shear banding is observed for $\dot \gamma > \dot \gamma^*$. Such a behaviour can be interpreted in the framework recently described by Martin and Hu on Laponite samples \cite{Martin:2012}. Under an external shear, the sample in the vicinity of the rotor that is in initially at rest gets fluidized, while the sample close to the stator remains at rest and keeps aging. Depending on the intensity of the shear rate, such heterogeneous velocity profile may either be trapped by the sample aging and become permanent as observed below $\dot \gamma ^*$, or become homogeneous after a transient shear-banding regime which duration decreases for increasing values of $\dot \gamma$. Such an argument allows us to understand that $\dot \gamma ^*$ will be sensitive to the gap size and to the cell geometry, which we have verified by performing identical shear startup experiments with different gap widths. These experiments show that $\dot \gamma^*$ decreases as the gap size is reduced [see Fig.~6 in the ESI]. Therefore, steady-state shear banding trapped by aging should be distinguished from shear banding resulting from intrinsic flow instability which should not depend on the geometry. Finally, we have shown in section~\ref{PD} that $\dot \gamma ^*$ is a nonlinear function of the sample age $t_w$, again in contrast with $\dot \gamma_c$.

Regarding the transient fluidization dynamics, our work unravels the existence of very characteristic peaks in the global rheological data. These peaks correspond to local avalanches associated with the abrupt fluidization of shear-banded velocity profiles. An avalanche proceeds in two steps. First, the sample ages as evidenced by the slow increase of the stress indicative of progressive consolidation, while the shear band remains roughly fixed. Second, the sample suddenly fluidizes before localizing again, at least partially, while the stress drops and the shear rate increases. This scenario strikingly recalls the transient shear banding reported in Laponite clay suspensions \cite{Martin:2012} and fits well with the stability criterion recently proposed by Fielding {\it et al.} \cite{Moorcroft:2013,Fielding:2014}. In practice though, the physical origin of the avalanche-like and successive stress relaxation events remains to be determined. Aging is more pronounced in our system than in the attractive gels that have been studied previously \cite{Derec:2003,Coussot:2006,Ovarlez:2007} as evidenced from the large values of $\textrm{d} G'/\textrm{d} \log t$ after preshear [see inset of Fig.~\ref{fig2}(a)]. Therefore, syneresis driven by the aggregation and/or sedimentation of the colloidal flocs due to their density mismatch with the surrounding fluid could also play a role and interfere with the traditional picture of a competition between aging and shear rejuvenation \cite{Moller:2006}. As syneresis is negligible in most of the colloidal gels that have been the topic of rheophysical studies so far, it could also explain why such stress oscillations have, to our knowledge, never been reported in the literature.

An alternative explanation for the stress oscillations could be related to confinement. Indeed, as the size distribution of the fused silica aggregates is wide and extends up to 100~$\mu$m, the sample mesostructure most likely involves aggregates which size becomes comparable to that of the gap, at least for long enough rest durations $t_w$. In this framework, stress oscillations would result from a competition between shear-induced structuration as described in the introduction and the strong aging of the sample. Such an interpretation would also account for the stick-slip like motion of the fluidized band at the moving wall. Nonetheless, despite systematic monitoring of the sample during shear startup experiments, no structuration or spatial pattern could be observed. Moreover, supplemental shear startup experiments in narrower gaps show that for a given shear rate, stronger confinement leads to the disappearance of the stress oscillations and to homogeneous velocity profiles [see Fig.~6 in the ESI]. This last result strongly suggests that confinement alone cannot account for stress oscillations. 

Finally, the present study has focused on experiments performed under smooth boundary conditions, revealing the presence of strong wall slip associated with heterogeneous, shear-banded flows while fully fluidized states show negligible wall slip. Yet the roughness and/or chemical nature of the walls are known to have a crucial influence not only on rheological measurements but also on local flow both close to the walls and in the bulk \cite{Gibaud:2008,Gibaud:2009,Seth:2012,Grenard:2014,Mansard:2014}. Therefore the influence of boundary conditions on the complex fluidization scenario reported here appears as the next key question to address in future work. As a first step, we report preliminary tests on the role of boundary conditions on the above results in the ESI. The flow curve measured under ``rough'' boundary conditions in a sand-blasted Plexiglas Taylor-Couette cell with typical roughness of 1~$\mu$m (to be compared to a few tens of nanometers for the ``smooth'' cell used so far) shows a smaller, yet significant hysteresis [see Fig.~1(b) in the ESI] as well as wall slip at low shear rates (see Fig.~5 in the ESI). 
Furthermore, one can see a strong difference between the velocity profiles recorded simultaneously to the flow curve respectively with the rough and smooth boundary conditions (compare Fig.~4 and 5 in the ESI) although the surface roughness of the rough boundary does not match exactly the size of the microstructure. In particular the transient fluidization episode reported around 7~s$^{-1}$ with smooth boundary conditions [Fig.~\ref{fig2}(b) and Fig.~4 in the ESI] is absent with rough walls (see Fig.~5 in the ESI). These preliminary results illustrate the strong impact of the boundary conditions and urge for systematic experiments so as to quantify the impact of the boundary conditions especially on the state diagram reported in Fig.~\ref{fig4}.

\section{Conclusion}

We have investigated the local scenario associated with the shear-induced fluidization of an attractive gel made of non-Brownian particles. We have identified a critical shear rate $\dot \gamma^*$ that separates steady shear-banded flows from full fluidization and that exhibits a nonlinear dependence with sample age $t_w$. This critical shear rate $\dot \gamma^*$ is much larger that the one signaling flow instability in experiments starting from fluidized states and  depends on the shear cell geometry as well as possibly on the preshear intensity imposed prior to the shear startup experiment. As a key result, we have shown that for shear rates larger than $\dot \gamma^*$, the fluidization of the sample involves successive local avalanche-like events that are heterogeneously distributed along the cell height and which dynamics is strongly coupled to both the slip behaviour at the wall and the width of the shear band. Such avalanches appear in the stress signal as peaks, which individual properties are strongly reminiscent of stick-slip phenomena. Future work will focus on the early stage of shear startup experiments, and in particular on the stress overshoot that occurs before the stress relaxation, as well as on the influence of the boundaries and confinement on the fluidization scenario.
    
\begin{acknowledgments}    
The authors thank J.-P.~Chapel, B.~Keshavarz and G.~Ovarlez for stimulating discussions, T.~Gibaud for allowing us to use his microscope, J.~Laurent for her precious assistance with the SEM experiments as well as two anonymous referees for their constructive remarks on our manuscript. This work was supported by JSPS and CNRS under the Japan-France Research Cooperative Program (PRC CNRS/JSPS RheoVolc). A.K. acknowledges support by JSPS Kakenhi Grant No. 265204. S.M.~acknowledges funding from Institut Universitaire de France and from the European Research Council under the European Union's Seventh Framework Programme (FP7/2007-2013) and ERC Grant Agreement No.~258803.
    
\end{acknowledgments}


\begin{thebibliography}{74}%
\providecommand*{\natexlab}[1]{#1}
\providecommand*{\mciteSetBstSublistMode}[1]{}
\providecommand*{\mciteSetBstMaxWidthForm}[2]{}
\providecommand*{\mciteBstWouldAddEndPuncttrue}
  {\def\EndOfBibitem{\unskip.}}
\providecommand*{\mciteBstWouldAddEndPunctfalse}
  {\let\EndOfBibitem\relax}
\providecommand*{\mciteSetBstMidEndSepPunct}[3]{}
\providecommand*{\mciteSetBstSublistLabelBeginEnd}[3]{}
\providecommand*{\EndOfBibitem}{}
\mciteSetBstSublistMode{f}
\mciteSetBstMaxWidthForm{subitem}
{(\emph{\alph{mcitesubitemcount}})}
\mciteSetBstSublistLabelBeginEnd{\mcitemaxwidthsubitemform\space}
{\relax}{\relax}

\bibitem[Coussot(2015)]{Coussot:2015}
P.~Coussot, \emph{J. Non-Newtonian Fluid Mech.}, 2015, \textbf{211},
  31--49\relax
\mciteBstWouldAddEndPuncttrue
\mciteSetBstMidEndSepPunct{\mcitedefaultmidpunct}
{\mcitedefaultendpunct}{\mcitedefaultseppunct}\relax
\EndOfBibitem
\bibitem[Bonn \emph{et~al.}(2015)Bonn, Paredes, Denn, Berthier, Divoux, and
  Manneville]{Bonn:2015}
D.~Bonn, J.~Paredes, M.~Denn, L.~Berthier, T.~Divoux and S.~Manneville,
  \emph{arXiv:1502.05281}, 2015\relax
\mciteBstWouldAddEndPuncttrue
\mciteSetBstMidEndSepPunct{\mcitedefaultmidpunct}
{\mcitedefaultendpunct}{\mcitedefaultseppunct}\relax
\EndOfBibitem
\bibitem[M{\o}ller \emph{et~al.}(2009)M{\o}ller, Fall, and Bonn]{Moller:2009a}
P.~C.~F. M{\o}ller, A.~Fall and D.~Bonn, \emph{Europhys. Lett.}, 2009,
  \textbf{87}, 38004\relax
\mciteBstWouldAddEndPuncttrue
\mciteSetBstMidEndSepPunct{\mcitedefaultmidpunct}
{\mcitedefaultendpunct}{\mcitedefaultseppunct}\relax
\EndOfBibitem
\bibitem[Ovarlez \emph{et~al.}(2013)Ovarlez, Cohen-Addad, Krishan, Goyon, and
  Coussot]{Ovarlez:2013b}
G.~Ovarlez, S.~Cohen-Addad, K.~Krishan, J.~Goyon and P.~Coussot, \emph{J.
  Non-Newtonian Fluid Mech.}, 2013, \textbf{193}, 68--79\relax
\mciteBstWouldAddEndPuncttrue
\mciteSetBstMidEndSepPunct{\mcitedefaultmidpunct}
{\mcitedefaultendpunct}{\mcitedefaultseppunct}\relax
\EndOfBibitem
\bibitem[Balmforth \emph{et~al.}(2014)Balmforth, Frigaard, and
  Ovarlez]{Balmforth:2014}
N.~Balmforth, I.~Frigaard and G.~Ovarlez, \emph{Annu. Rev. Fluid Mech.}, 2014,
  \textbf{46}, 121--146\relax
\mciteBstWouldAddEndPuncttrue
\mciteSetBstMidEndSepPunct{\mcitedefaultmidpunct}
{\mcitedefaultendpunct}{\mcitedefaultseppunct}\relax
\EndOfBibitem
\bibitem[Bonnecaze and Cloitre(2010)]{Bonnecaze:2010}
R.~Bonnecaze and M.~Cloitre, \emph{Adv. Polym. Sci.}, 2010, \textbf{236},
  117--161\relax
\mciteBstWouldAddEndPuncttrue
\mciteSetBstMidEndSepPunct{\mcitedefaultmidpunct}
{\mcitedefaultendpunct}{\mcitedefaultseppunct}\relax
\EndOfBibitem
\bibitem[Mason \emph{et~al.}(1996)Mason, Bibette, and Weitz]{Mason:1996a}
T.~G. Mason, J.~Bibette and D.~A. Weitz, \emph{J. Colloid Interface Sci.},
  1996, \textbf{179}, 439--448\relax
\mciteBstWouldAddEndPuncttrue
\mciteSetBstMidEndSepPunct{\mcitedefaultmidpunct}
{\mcitedefaultendpunct}{\mcitedefaultseppunct}\relax
\EndOfBibitem
\bibitem[Cloitre \emph{et~al.}(2003)Cloitre, Borrega, Monti, and
  Leibler]{Cloitre:2003a}
M.~Cloitre, M.~Borrega, F.~Monti and L.~Leibler, \emph{C.R. Physique}, 2003,
  \textbf{4}, 221--230\relax
\mciteBstWouldAddEndPuncttrue
\mciteSetBstMidEndSepPunct{\mcitedefaultmidpunct}
{\mcitedefaultendpunct}{\mcitedefaultseppunct}\relax
\EndOfBibitem
\bibitem[Seth \emph{et~al.}(2006)Seth, Cloitre, and Bonnecaze]{Seth:2006}
J.~Seth, M.~Cloitre and R.~Bonnecaze, \emph{J. Rheol.}, 2006, \textbf{50},
  353--376\relax
\mciteBstWouldAddEndPuncttrue
\mciteSetBstMidEndSepPunct{\mcitedefaultmidpunct}
{\mcitedefaultendpunct}{\mcitedefaultseppunct}\relax
\EndOfBibitem
\bibitem[Siebenb{\"u}rger \emph{et~al.}(2012)Siebenb{\"u}rger, Fuchs, and
  Ballauff]{Siebenburger:2012b}
M.~Siebenb{\"u}rger, M.~Fuchs and M.~Ballauff, \emph{Soft Matter}, 2012,
  \textbf{8}, 4014--4024\relax
\mciteBstWouldAddEndPuncttrue
\mciteSetBstMidEndSepPunct{\mcitedefaultmidpunct}
{\mcitedefaultendpunct}{\mcitedefaultseppunct}\relax
\EndOfBibitem
\bibitem[Lu \emph{et~al.}(2008)Lu, Zaccarelli, Ciulla, Schofield, Sciortino,
  and Weitz]{Lu:2008}
P.~Lu, E.~Zaccarelli, F.~Ciulla, A.~B. Schofield, F.~Sciortino and D.~A. Weitz,
  \emph{Nature}, 2008, \textbf{453}, 499--503\relax
\mciteBstWouldAddEndPuncttrue
\mciteSetBstMidEndSepPunct{\mcitedefaultmidpunct}
{\mcitedefaultendpunct}{\mcitedefaultseppunct}\relax
\EndOfBibitem
\bibitem[Lu and Weitz(2013)]{Lu:2013}
P.~Lu and D.~Weitz, \emph{Annu. Rev. Condens. Matter Phys.}, 2013, \textbf{4},
  217--233\relax
\mciteBstWouldAddEndPuncttrue
\mciteSetBstMidEndSepPunct{\mcitedefaultmidpunct}
{\mcitedefaultendpunct}{\mcitedefaultseppunct}\relax
\EndOfBibitem
\bibitem[Mourchid \emph{et~al.}(1995)Mourchid, Delville, Lambard, L{\'e}colier,
  and Levitz]{Mourchid:1995}
A.~Mourchid, A.~Delville, J.~Lambard, E.~L{\'e}colier and P.~Levitz,
  \emph{Langmuir}, 1995, \textbf{11}, 1942--1950\relax
\mciteBstWouldAddEndPuncttrue
\mciteSetBstMidEndSepPunct{\mcitedefaultmidpunct}
{\mcitedefaultendpunct}{\mcitedefaultseppunct}\relax
\EndOfBibitem
\bibitem[Pignon \emph{et~al.}(1997)Pignon, Magnin, Piau, Cabane, Lindner, and
  Diat]{Pignon:1997b}
F.~Pignon, A.~Magnin, J.-M. Piau, B.~Cabane, P.~Lindner and O.~Diat,
  \emph{Phys. Rev. E}, 1997, \textbf{56}, 3281--3289\relax
\mciteBstWouldAddEndPuncttrue
\mciteSetBstMidEndSepPunct{\mcitedefaultmidpunct}
{\mcitedefaultendpunct}{\mcitedefaultseppunct}\relax
\EndOfBibitem
\bibitem[Paineau \emph{et~al.}(2011)Paineau, Bihannic, Baravian, Philippe,
  Davidson, Levitz, Funari, Rochas, and Michot]{Paineau:2011}
E.~Paineau, I.~Bihannic, C.~Baravian, A.-M. Philippe, P.~Davidson, P.~Levitz,
  S.~Funari, C.~Rochas and L.~Michot, \emph{Langmuir}, 2011, \textbf{27},
  5562--5573\relax
\mciteBstWouldAddEndPuncttrue
\mciteSetBstMidEndSepPunct{\mcitedefaultmidpunct}
{\mcitedefaultendpunct}{\mcitedefaultseppunct}\relax
\EndOfBibitem
\bibitem[Trappe and Weitz(2000)]{Trappe:2000}
V.~Trappe and D.~A. Weitz, \emph{Phys. Rev. Lett.}, 2000, \textbf{85},
  449--452\relax
\mciteBstWouldAddEndPuncttrue
\mciteSetBstMidEndSepPunct{\mcitedefaultmidpunct}
{\mcitedefaultendpunct}{\mcitedefaultseppunct}\relax
\EndOfBibitem
\bibitem[Trappe \emph{et~al.}(2001)Trappe, Prasad, Cipelletti, Segre, and
  Weitz]{Trappe:2001}
V.~Trappe, V.~Prasad, L.~Cipelletti, P.~N. Segre and D.~A. Weitz,
  \emph{Nature}, 2001, \textbf{411}, 772--775\relax
\mciteBstWouldAddEndPuncttrue
\mciteSetBstMidEndSepPunct{\mcitedefaultmidpunct}
{\mcitedefaultendpunct}{\mcitedefaultseppunct}\relax
\EndOfBibitem
\bibitem[Pham \emph{et~al.}(2004)Pham, Egelhaaf, Pusey, and Poon]{Pham:2004}
K.~Pham, S.~Egelhaaf, P.~Pusey and W.~Poon, \emph{Phys. Rev. E}, 2004,
  \textbf{69}, 011503\relax
\mciteBstWouldAddEndPuncttrue
\mciteSetBstMidEndSepPunct{\mcitedefaultmidpunct}
{\mcitedefaultendpunct}{\mcitedefaultseppunct}\relax
\EndOfBibitem
\bibitem[Mock and Zukoski(2007)]{Mock:2007}
E.~Mock and C.~Zukoski, \emph{J. Rheol.}, 2007, \textbf{51}, 541--559\relax
\mciteBstWouldAddEndPuncttrue
\mciteSetBstMidEndSepPunct{\mcitedefaultmidpunct}
{\mcitedefaultendpunct}{\mcitedefaultseppunct}\relax
\EndOfBibitem
\bibitem[Reddy \emph{et~al.}(2012)Reddy, Zhang, Lettinga, Dhont, and
  Vermant]{Reddy:2012}
N.~Reddy, Z.~Zhang, M.~Lettinga, J.~Dhont and J.~Vermant, \emph{J. Rheol.},
  2012, \textbf{56}, 1153--1174\relax
\mciteBstWouldAddEndPuncttrue
\mciteSetBstMidEndSepPunct{\mcitedefaultmidpunct}
{\mcitedefaultendpunct}{\mcitedefaultseppunct}\relax
\EndOfBibitem
\bibitem[Cipelletti \emph{et~al.}(2000)Cipelletti, Manley, Ball, and
  Weitz]{Cipelletti:2000}
L.~Cipelletti, S.~Manley, R.~C. Ball and D.~A. Weitz, \emph{Phys. Rev. Lett.},
  2000, \textbf{84}, 2275--2278\relax
\mciteBstWouldAddEndPuncttrue
\mciteSetBstMidEndSepPunct{\mcitedefaultmidpunct}
{\mcitedefaultendpunct}{\mcitedefaultseppunct}\relax
\EndOfBibitem
\bibitem[Derec \emph{et~al.}(2003)Derec, Ducouret, Ajdari, and
  Lequeux]{Derec:2003}
C.~Derec, G.~Ducouret, A.~Ajdari and F.~Lequeux, \emph{Phys. Rev. E}, 2003,
  \textbf{67}, 061403\relax
\mciteBstWouldAddEndPuncttrue
\mciteSetBstMidEndSepPunct{\mcitedefaultmidpunct}
{\mcitedefaultendpunct}{\mcitedefaultseppunct}\relax
\EndOfBibitem
\bibitem[Ovarlez and Chateau(2008)]{Ovarlez:2008b}
G.~Ovarlez and X.~Chateau, \emph{Phys. Rev. E}, 2008, \textbf{77}, 061403\relax
\mciteBstWouldAddEndPuncttrue
\mciteSetBstMidEndSepPunct{\mcitedefaultmidpunct}
{\mcitedefaultendpunct}{\mcitedefaultseppunct}\relax
\EndOfBibitem
\bibitem[Negi and Osuji(2009)]{Negi:2009b}
A.~Negi and C.~Osuji, \emph{Phys. Rev. E}, 2009, \textbf{80}, 010404\relax
\mciteBstWouldAddEndPuncttrue
\mciteSetBstMidEndSepPunct{\mcitedefaultmidpunct}
{\mcitedefaultendpunct}{\mcitedefaultseppunct}\relax
\EndOfBibitem
\bibitem[Koumakis and Petekidis(2011)]{Koumakis:2011}
N.~Koumakis and G.~Petekidis, \emph{Soft Matter}, 2011, \textbf{7},
  2456--2470\relax
\mciteBstWouldAddEndPuncttrue
\mciteSetBstMidEndSepPunct{\mcitedefaultmidpunct}
{\mcitedefaultendpunct}{\mcitedefaultseppunct}\relax
\EndOfBibitem
\bibitem[Scherer(1989)]{Scherer:1989}
G.~Scherer, \emph{J. non-Cryst. Solids}, 1989, \textbf{108}, 18--27\relax
\mciteBstWouldAddEndPuncttrue
\mciteSetBstMidEndSepPunct{\mcitedefaultmidpunct}
{\mcitedefaultendpunct}{\mcitedefaultseppunct}\relax
\EndOfBibitem
\bibitem[Manley \emph{et~al.}(2005)Manley, Skotheim, Mahadevan, and
  Weitz]{Manley:2005}
S.~Manley, J.~Skotheim, L.~Mahadevan and D.~Weitz, \emph{Phys. Rev. Lett.},
  2005, \textbf{94}, 218302\relax
\mciteBstWouldAddEndPuncttrue
\mciteSetBstMidEndSepPunct{\mcitedefaultmidpunct}
{\mcitedefaultendpunct}{\mcitedefaultseppunct}\relax
\EndOfBibitem
\bibitem[Buscall \emph{et~al.}(2009)Buscall, Choudhury, Faers, Goodwin,
  Luckham, and Partridge.]{Buscall:2009}
R.~Buscall, T.~H. Choudhury, M.~A. Faers, J.~W. Goodwin, P.~A. Luckham and
  S.~J. Partridge., \emph{Soft Matter}, 2009, \textbf{5}, 1345--1349\relax
\mciteBstWouldAddEndPuncttrue
\mciteSetBstMidEndSepPunct{\mcitedefaultmidpunct}
{\mcitedefaultendpunct}{\mcitedefaultseppunct}\relax
\EndOfBibitem
\bibitem[Brambilla \emph{et~al.}(2011)Brambilla, Buzzaccaro, Piazza, Berthier,
  and Cipelletti]{Brambilla:2011}
G.~Brambilla, S.~Buzzaccaro, R.~Piazza, L.~Berthier and L.~Cipelletti,
  \emph{Phys. Rev. Lett.}, 2011, \textbf{106}, 118302\relax
\mciteBstWouldAddEndPuncttrue
\mciteSetBstMidEndSepPunct{\mcitedefaultmidpunct}
{\mcitedefaultendpunct}{\mcitedefaultseppunct}\relax
\EndOfBibitem
\bibitem[Teece \emph{et~al.}(2011)Teece, Faers, and Bartlett]{Teece:2011}
L.~J. Teece, M.~A. Faers and P.~Bartlett, \emph{Soft Matter}, 2011, \textbf{7},
  1341--1351\relax
\mciteBstWouldAddEndPuncttrue
\mciteSetBstMidEndSepPunct{\mcitedefaultmidpunct}
{\mcitedefaultendpunct}{\mcitedefaultseppunct}\relax
\EndOfBibitem
\bibitem[Barlett \emph{et~al.}(2012)Barlett, Teece, and Faers]{Barlett:2012}
P.~Barlett, L.~Teece and M.~Faers, \emph{Phys. Rev. E}, 2012, \textbf{85},
  021404\relax
\mciteBstWouldAddEndPuncttrue
\mciteSetBstMidEndSepPunct{\mcitedefaultmidpunct}
{\mcitedefaultendpunct}{\mcitedefaultseppunct}\relax
\EndOfBibitem
\bibitem[Buscall and McGowan(1983)]{Buscall:1983}
R.~Buscall and I.~McGowan, \emph{Faraday Discuss. Chem. Soc.}, 1983,
  \textbf{76}, 277--290\relax
\mciteBstWouldAddEndPuncttrue
\mciteSetBstMidEndSepPunct{\mcitedefaultmidpunct}
{\mcitedefaultendpunct}{\mcitedefaultseppunct}\relax
\EndOfBibitem
\bibitem[Gibaud \emph{et~al.}(2008)Gibaud, Barentin, and
  Manneville]{Gibaud:2008}
T.~Gibaud, C.~Barentin and S.~Manneville, \emph{Phys. Rev. Lett.}, 2008,
  \textbf{101}, 258302\relax
\mciteBstWouldAddEndPuncttrue
\mciteSetBstMidEndSepPunct{\mcitedefaultmidpunct}
{\mcitedefaultendpunct}{\mcitedefaultseppunct}\relax
\EndOfBibitem
\bibitem[Gibaud \emph{et~al.}(2009)Gibaud, Barentin, Taberlet, and
  Manneville]{Gibaud:2009}
T.~Gibaud, C.~Barentin, N.~Taberlet and S.~Manneville, \emph{Soft Matter},
  2009, \textbf{5}, 3026--3037\relax
\mciteBstWouldAddEndPuncttrue
\mciteSetBstMidEndSepPunct{\mcitedefaultmidpunct}
{\mcitedefaultendpunct}{\mcitedefaultseppunct}\relax
\EndOfBibitem
\bibitem[DeGroot \emph{et~al.}(1994)DeGroot, Macosko, Kume, and
  Hashimoto]{DeGroot:1994}
J.~J.~V. DeGroot, C.~W. Macosko, T.~Kume and T.~Hashimoto, \emph{J. Colloid
  Interface Sci.}, 1994, \textbf{166}, 404--413\relax
\mciteBstWouldAddEndPuncttrue
\mciteSetBstMidEndSepPunct{\mcitedefaultmidpunct}
{\mcitedefaultendpunct}{\mcitedefaultseppunct}\relax
\EndOfBibitem
\bibitem[Navarrete \emph{et~al.}(1996)Navarrete, Scriven, and
  Macosko]{Navarrete:1996}
R.~Navarrete, L.~Scriven and C.~Macosko, \emph{J. Colloid Interface Sci.},
  1996, \textbf{180}, 200--211\relax
\mciteBstWouldAddEndPuncttrue
\mciteSetBstMidEndSepPunct{\mcitedefaultmidpunct}
{\mcitedefaultendpunct}{\mcitedefaultseppunct}\relax
\EndOfBibitem
\bibitem[Montesi \emph{et~al.}(2004)Montesi, Pe${\rm \tilde{n}}$a, and
  Pasquali]{Montesi:2004}
A.~Montesi, A.~Pe${\rm \tilde{n}}$a and M.~Pasquali, \emph{Phys. Rev. Lett.},
  2004, \textbf{92}, 058303\relax
\mciteBstWouldAddEndPuncttrue
\mciteSetBstMidEndSepPunct{\mcitedefaultmidpunct}
{\mcitedefaultendpunct}{\mcitedefaultseppunct}\relax
\EndOfBibitem
\bibitem[Osuji \emph{et~al.}(2008)Osuji, Kim, and Weitz]{Osuji:2008}
C.~O. Osuji, C.~Kim and D.~A. Weitz, \emph{Phys. Rev. E}, 2008, \textbf{77},
  060402(R)\relax
\mciteBstWouldAddEndPuncttrue
\mciteSetBstMidEndSepPunct{\mcitedefaultmidpunct}
{\mcitedefaultendpunct}{\mcitedefaultseppunct}\relax
\EndOfBibitem
\bibitem[Negi and Osuji(2009)]{Negi:2009}
A.~Negi and C.~Osuji, \emph{Rheol. Acta}, 2009, \textbf{48}, 871--881\relax
\mciteBstWouldAddEndPuncttrue
\mciteSetBstMidEndSepPunct{\mcitedefaultmidpunct}
{\mcitedefaultendpunct}{\mcitedefaultseppunct}\relax
\EndOfBibitem
\bibitem[Grenard \emph{et~al.}(2011)Grenard, Taberlet, and
  Manneville]{Grenard:2011}
V.~Grenard, N.~Taberlet and S.~Manneville, \emph{Soft Matter}, 2011,
  \textbf{7}, 3920--3928\relax
\mciteBstWouldAddEndPuncttrue
\mciteSetBstMidEndSepPunct{\mcitedefaultmidpunct}
{\mcitedefaultendpunct}{\mcitedefaultseppunct}\relax
\EndOfBibitem
\bibitem[Vermant and Solomon(2005)]{Vermant:2005}
J.~Vermant and M.~J. Solomon, \emph{J. Phys.: Condens. Matter}, 2005,
  \textbf{17}, R187--R216\relax
\mciteBstWouldAddEndPuncttrue
\mciteSetBstMidEndSepPunct{\mcitedefaultmidpunct}
{\mcitedefaultendpunct}{\mcitedefaultseppunct}\relax
\EndOfBibitem
\bibitem[Gopalakrishnan and Zukoski(2007)]{Gopalakrishnan:2007}
V.~Gopalakrishnan and C.~Zukoski, \emph{J. Rheol.}, 2007, \textbf{51},
  623--644\relax
\mciteBstWouldAddEndPuncttrue
\mciteSetBstMidEndSepPunct{\mcitedefaultmidpunct}
{\mcitedefaultendpunct}{\mcitedefaultseppunct}\relax
\EndOfBibitem
\bibitem[Gibaud \emph{et~al.}(2010)Gibaud, Frelat, and Manneville]{Gibaud:2010}
T.~Gibaud, D.~Frelat and S.~Manneville, \emph{Soft Matter}, 2010, \textbf{6},
  3482--3488\relax
\mciteBstWouldAddEndPuncttrue
\mciteSetBstMidEndSepPunct{\mcitedefaultmidpunct}
{\mcitedefaultendpunct}{\mcitedefaultseppunct}\relax
\EndOfBibitem
\bibitem[Sprakel \emph{et~al.}(2011)Sprakel, Lindstr\"om, Kodger, and
  Weitz]{Sprakel:2011}
J.~Sprakel, S.~Lindstr\"om, T.~Kodger and D.~Weitz, \emph{Phys. Rev. Lett.},
  2011, \textbf{106}, 248303\relax
\mciteBstWouldAddEndPuncttrue
\mciteSetBstMidEndSepPunct{\mcitedefaultmidpunct}
{\mcitedefaultendpunct}{\mcitedefaultseppunct}\relax
\EndOfBibitem
\bibitem[Grenard \emph{et~al.}(2014)Grenard, Divoux, Taberlet, and
  Manneville]{Grenard:2014}
V.~Grenard, T.~Divoux, N.~Taberlet and S.~Manneville, \emph{Soft Matter}, 2014,
  \textbf{10}, 1555--1571\relax
\mciteBstWouldAddEndPuncttrue
\mciteSetBstMidEndSepPunct{\mcitedefaultmidpunct}
{\mcitedefaultendpunct}{\mcitedefaultseppunct}\relax
\EndOfBibitem
\bibitem[Stickland \emph{et~al.}(2015)Stickland, Kumar, Kusuma, Scales,
  Tindley, Biggs, and Buscall]{Stickland:2015}
A.~Stickland, A.~Kumar, T.~Kusuma, P.~Scales, A.~Tindley, S.~Biggs and
  R.~Buscall, \emph{Rheol. Acta}, 2015, \textbf{54}, 337--352\relax
\mciteBstWouldAddEndPuncttrue
\mciteSetBstMidEndSepPunct{\mcitedefaultmidpunct}
{\mcitedefaultendpunct}{\mcitedefaultseppunct}\relax
\EndOfBibitem
\bibitem[Chan and Mohraz(2012)]{Chan:2012}
H.~Chan and A.~Mohraz, \emph{Phys. Rev. E}, 2012, \textbf{85}, 041403\relax
\mciteBstWouldAddEndPuncttrue
\mciteSetBstMidEndSepPunct{\mcitedefaultmidpunct}
{\mcitedefaultendpunct}{\mcitedefaultseppunct}\relax
\EndOfBibitem
\bibitem[Perge \emph{et~al.}(2014)Perge, Taberlet, Gibaud, and
  Manneville]{Perge:2014b}
C.~Perge, N.~Taberlet, T.~Gibaud and S.~Manneville, \emph{J. Rheol.}, 2014,
  \textbf{58}, 1331--1357\relax
\mciteBstWouldAddEndPuncttrue
\mciteSetBstMidEndSepPunct{\mcitedefaultmidpunct}
{\mcitedefaultendpunct}{\mcitedefaultseppunct}\relax
\EndOfBibitem
\bibitem[Varadan and Solomon(2001)]{Varadan:2001}
P.~Varadan and M.~Solomon, \emph{Langmuir}, 2001, \textbf{17}, 2918--2929\relax
\mciteBstWouldAddEndPuncttrue
\mciteSetBstMidEndSepPunct{\mcitedefaultmidpunct}
{\mcitedefaultendpunct}{\mcitedefaultseppunct}\relax
\EndOfBibitem
\bibitem[Mohraz and Solomon(2005)]{Mohraz:2005}
A.~Mohraz and M.~Solomon, \emph{J. Rheol.}, 2005, \textbf{49}, 657--681\relax
\mciteBstWouldAddEndPuncttrue
\mciteSetBstMidEndSepPunct{\mcitedefaultmidpunct}
{\mcitedefaultendpunct}{\mcitedefaultseppunct}\relax
\EndOfBibitem
\bibitem[Rajaram and Mohraz(2010)]{Rajaram:2010}
B.~Rajaram and A.~Mohraz, \emph{Soft Matter}, 2010, \textbf{6},
  2246--2259\relax
\mciteBstWouldAddEndPuncttrue
\mciteSetBstMidEndSepPunct{\mcitedefaultmidpunct}
{\mcitedefaultendpunct}{\mcitedefaultseppunct}\relax
\EndOfBibitem
\bibitem[Divoux \emph{et~al.}(2016)Divoux, Fardin, Manneville, and
  Lerouge]{Divoux:2016}
T.~Divoux, M.-A. Fardin, S.~Manneville and S.~Lerouge, \emph{Annu. Rev. Fluid
  Mech.}, 2016, \textbf{48}, 81--103\relax
\mciteBstWouldAddEndPuncttrue
\mciteSetBstMidEndSepPunct{\mcitedefaultmidpunct}
{\mcitedefaultendpunct}{\mcitedefaultseppunct}\relax
\EndOfBibitem
\bibitem[Raynaud \emph{et~al.}(2002)Raynaud, Moucheront, Baudez, Bertrand,
  Guilbaud, and Coussot]{Raynaud:2002}
J.~S. Raynaud, P.~Moucheront, J.~C. Baudez, F.~Bertrand, J.~P. Guilbaud and
  P.~Coussot, \emph{J. Rheol.}, 2002, \textbf{46}, 709--732\relax
\mciteBstWouldAddEndPuncttrue
\mciteSetBstMidEndSepPunct{\mcitedefaultmidpunct}
{\mcitedefaultendpunct}{\mcitedefaultseppunct}\relax
\EndOfBibitem
\bibitem[Martin and Hu(2012)]{Martin:2012}
J.~Martin and Y.~Hu, \emph{Soft Matter}, 2012, \textbf{8}, 6940--6949\relax
\mciteBstWouldAddEndPuncttrue
\mciteSetBstMidEndSepPunct{\mcitedefaultmidpunct}
{\mcitedefaultendpunct}{\mcitedefaultseppunct}\relax
\EndOfBibitem
\bibitem[Ovarlez \emph{et~al.}(2009)Ovarlez, Rodts, Chateau, and
  Coussot]{Ovarlez:2009}
G.~Ovarlez, S.~Rodts, X.~Chateau and P.~Coussot, \emph{Rheol. Acta}, 2009,
  \textbf{48}, 831--844\relax
\mciteBstWouldAddEndPuncttrue
\mciteSetBstMidEndSepPunct{\mcitedefaultmidpunct}
{\mcitedefaultendpunct}{\mcitedefaultseppunct}\relax
\EndOfBibitem
\bibitem[Fall \emph{et~al.}(2010)Fall, Paredes, and Bonn]{Fall:2010b}
A.~Fall, J.~Paredes and D.~Bonn, \emph{Phys. Rev. Lett.}, 2010, \textbf{105},
  225502\relax
\mciteBstWouldAddEndPuncttrue
\mciteSetBstMidEndSepPunct{\mcitedefaultmidpunct}
{\mcitedefaultendpunct}{\mcitedefaultseppunct}\relax
\EndOfBibitem
\bibitem[Bilmes(1942)]{Bilmes:1942}
L.~Bilmes, \emph{Nature}, 1942, \textbf{150}, 432--433\relax
\mciteBstWouldAddEndPuncttrue
\mciteSetBstMidEndSepPunct{\mcitedefaultmidpunct}
{\mcitedefaultendpunct}{\mcitedefaultseppunct}\relax
\EndOfBibitem
\bibitem[McKinley(2015)]{McKinley:2015}
G.~McKinley, \emph{Rheology Bulletin}, 2015, \textbf{84}, 14--17\relax
\mciteBstWouldAddEndPuncttrue
\mciteSetBstMidEndSepPunct{\mcitedefaultmidpunct}
{\mcitedefaultendpunct}{\mcitedefaultseppunct}\relax
\EndOfBibitem
\bibitem[Moorcroft and Fielding(2013)]{Moorcroft:2013}
R.~Moorcroft and S.~Fielding, \emph{Phys. Rev. Lett.}, 2013, \textbf{110},
  086001\relax
\mciteBstWouldAddEndPuncttrue
\mciteSetBstMidEndSepPunct{\mcitedefaultmidpunct}
{\mcitedefaultendpunct}{\mcitedefaultseppunct}\relax
\EndOfBibitem
\bibitem[Colombo \emph{et~al.}(2013)Colombo, Widmer-Cooper, and {Del
  Gado}]{Colombo:2013}
J.~Colombo, A.~Widmer-Cooper and E.~{Del Gado}, \emph{Phys. Rev. Lett.}, 2013,
  \textbf{110}, 198301\relax
\mciteBstWouldAddEndPuncttrue
\mciteSetBstMidEndSepPunct{\mcitedefaultmidpunct}
{\mcitedefaultendpunct}{\mcitedefaultseppunct}\relax
\EndOfBibitem
\bibitem[Colombo and {Del Gado}(2014)]{Colombo:2014}
J.~Colombo and E.~{Del Gado}, \emph{J. Rheol.}, 2014, \textbf{58},
  1089--1116\relax
\mciteBstWouldAddEndPuncttrue
\mciteSetBstMidEndSepPunct{\mcitedefaultmidpunct}
{\mcitedefaultendpunct}{\mcitedefaultseppunct}\relax
\EndOfBibitem
\bibitem[Fielding(2014)]{Fielding:2014}
S.~Fielding, \emph{Rep. Prog. Phys.}, 2014, \textbf{77}, 102601\relax
\mciteBstWouldAddEndPuncttrue
\mciteSetBstMidEndSepPunct{\mcitedefaultmidpunct}
{\mcitedefaultendpunct}{\mcitedefaultseppunct}\relax
\EndOfBibitem
\bibitem[Manneville \emph{et~al.}(2004)Manneville, B{\'e}cu, and
  Colin]{Manneville:2004a}
S.~Manneville, L.~B{\'e}cu and A.~Colin, \emph{Eur. Phys. J. AP}, 2004,
  \textbf{28}, 361--373\relax
\mciteBstWouldAddEndPuncttrue
\mciteSetBstMidEndSepPunct{\mcitedefaultmidpunct}
{\mcitedefaultendpunct}{\mcitedefaultseppunct}\relax
\EndOfBibitem
\bibitem[Gallot \emph{et~al.}(2013)Gallot, Perge, Grenard, Fardin, Taberlet,
  and Manneville]{Gallot:2013}
T.~Gallot, C.~Perge, V.~Grenard, M.-A. Fardin, N.~Taberlet and S.~Manneville,
  \emph{Rev. Sci. Instrum.}, 2013, \textbf{84}, 045107\relax
\mciteBstWouldAddEndPuncttrue
\mciteSetBstMidEndSepPunct{\mcitedefaultmidpunct}
{\mcitedefaultendpunct}{\mcitedefaultseppunct}\relax
\EndOfBibitem
\bibitem[M{\o}ller \emph{et~al.}(2008)M{\o}ller, Rodts, Michels, and
  Bonn]{Moller:2008}
P.~C.~F. M{\o}ller, S.~Rodts, M.~A.~J. Michels and D.~Bonn, \emph{Phys. Rev.
  E}, 2008, \textbf{77}, 041507\relax
\mciteBstWouldAddEndPuncttrue
\mciteSetBstMidEndSepPunct{\mcitedefaultmidpunct}
{\mcitedefaultendpunct}{\mcitedefaultseppunct}\relax
\EndOfBibitem
\bibitem[Kobayashi \emph{et~al.}(2005)Kobayashi, Juillerat, Galletto, Bowen,
  and Borkovec]{Kobayashi:2005}
M.~Kobayashi, F.~Juillerat, P.~Galletto, P.~Bowen and M.~Borkovec,
  \emph{Langmuir}, 2005, \textbf{21}, 5761--5769\relax
\mciteBstWouldAddEndPuncttrue
\mciteSetBstMidEndSepPunct{\mcitedefaultmidpunct}
{\mcitedefaultendpunct}{\mcitedefaultseppunct}\relax
\EndOfBibitem
\bibitem[Heston \emph{et~al.}(1960)Heston, Iler, and Sears]{Heston:1960}
W.~Heston, R.~Iler and G.~Sears, \emph{J. Chem. Phys.}, 1960, \textbf{64},
  147--150\relax
\mciteBstWouldAddEndPuncttrue
\mciteSetBstMidEndSepPunct{\mcitedefaultmidpunct}
{\mcitedefaultendpunct}{\mcitedefaultseppunct}\relax
\EndOfBibitem
\bibitem[Trompette and Clifton(2004)]{Trompette:2004}
J.~Trompette and M.~Clifton, \emph{J. Colloid Interface Sci.}, 2004,
  \textbf{276}, 475--482\relax
\mciteBstWouldAddEndPuncttrue
\mciteSetBstMidEndSepPunct{\mcitedefaultmidpunct}
{\mcitedefaultendpunct}{\mcitedefaultseppunct}\relax
\EndOfBibitem
\bibitem[Cao \emph{et~al.}(2010)Cao, Cummins, and Morris]{Cao:2010}
X.~J. Cao, H.~Z. Cummins and J.~F. Morris, \emph{Soft Matter}, 2010,
  \textbf{6}, 5425--5433\relax
\mciteBstWouldAddEndPuncttrue
\mciteSetBstMidEndSepPunct{\mcitedefaultmidpunct}
{\mcitedefaultendpunct}{\mcitedefaultseppunct}\relax
\EndOfBibitem
\bibitem[Truzzolillo \emph{et~al.}(2014)Truzzolillo, Roger, Dupas, Mora, and
  Cipelletti]{Truzzolillo:2014pp}
D.~Truzzolillo, V.~Roger, C.~Dupas, S.~Mora and L.~Cipelletti, E-print
  cond-mat/1411.2265\relax
\mciteBstWouldAddEndPuncttrue
\mciteSetBstMidEndSepPunct{\mcitedefaultmidpunct}
{\mcitedefaultendpunct}{\mcitedefaultseppunct}\relax
\EndOfBibitem
\bibitem[Allen and Matijevi\'c(1969)]{Allen:1969}
L.~Allen and E.~Matijevi\'c, \emph{J. Colloid Interface Sci.}, 1969,
  \textbf{31}, 287--296\relax
\mciteBstWouldAddEndPuncttrue
\mciteSetBstMidEndSepPunct{\mcitedefaultmidpunct}
{\mcitedefaultendpunct}{\mcitedefaultseppunct}\relax
\EndOfBibitem
\bibitem[Allen and Matijevi\'c(1970)]{Allen:1970}
L.~Allen and E.~Matijevi\'c, \emph{J. Colloid Interface Sci.}, 1970,
  \textbf{33}, 420--429\relax
\mciteBstWouldAddEndPuncttrue
\mciteSetBstMidEndSepPunct{\mcitedefaultmidpunct}
{\mcitedefaultendpunct}{\mcitedefaultseppunct}\relax
\EndOfBibitem
\bibitem[Depasse and Watillon(1970)]{Depasse:1970}
J.~Depasse and A.~Watillon, \emph{J. Colloid Interface Sci.}, 1970,
  \textbf{33}, 430--438\relax
\mciteBstWouldAddEndPuncttrue
\mciteSetBstMidEndSepPunct{\mcitedefaultmidpunct}
{\mcitedefaultendpunct}{\mcitedefaultseppunct}\relax
\EndOfBibitem
\bibitem[Depasse(1997)]{Depasse:1997}
J.~Depasse, \emph{J. Colloid Interface Sci.}, 1997, \textbf{194},
  260--262\relax
\mciteBstWouldAddEndPuncttrue
\mciteSetBstMidEndSepPunct{\mcitedefaultmidpunct}
{\mcitedefaultendpunct}{\mcitedefaultseppunct}\relax
\EndOfBibitem
\bibitem[Drabarek \emph{et~al.}(2002)Drabarek, Bartlett, Hanley, Woolfrey, and
  Muzny]{Drabarek:2002}
E.~Drabarek, J.~Bartlett, H.~Hanley, J.~Woolfrey and C.~Muzny, \emph{Int. J.
  Thermophys.}, 2002, \textbf{23}, 145--160\relax
\mciteBstWouldAddEndPuncttrue
\mciteSetBstMidEndSepPunct{\mcitedefaultmidpunct}
{\mcitedefaultendpunct}{\mcitedefaultseppunct}\relax
\EndOfBibitem
\bibitem[Coussot \emph{et~al.}(2002)Coussot, Raynaud, Bertrand, Moucheront,
  Guilbaud, Huynh, Jarny, and Lesueur]{Coussot:2002a}
P.~Coussot, J.~S. Raynaud, F.~Bertrand, P.~Moucheront, J.~P. Guilbaud, H.~T.
  Huynh, S.~Jarny and D.~Lesueur, \emph{Phys. Rev. Lett.}, 2002, \textbf{88},
  218301\relax
\mciteBstWouldAddEndPuncttrue
\mciteSetBstMidEndSepPunct{\mcitedefaultmidpunct}
{\mcitedefaultendpunct}{\mcitedefaultseppunct}\relax
\EndOfBibitem
\bibitem[Ragouilliaux \emph{et~al.}(2006)Ragouilliaux, Herzhaft, Bertrand, and
  Coussot]{Ragouilliaux:2006}
A.~Ragouilliaux, B.~Herzhaft, F.~Bertrand and P.~Coussot, \emph{Rheol. Acta},
  2006, \textbf{46}, 261--271\relax
\mciteBstWouldAddEndPuncttrue
\mciteSetBstMidEndSepPunct{\mcitedefaultmidpunct}
{\mcitedefaultendpunct}{\mcitedefaultseppunct}\relax
\EndOfBibitem
\bibitem[ten Brinke \emph{et~al.}(2007)ten Brinke, Bailey, Lekkerkerker, and
  Maitland]{tenBrinke:2007}
A.~ten Brinke, L.~Bailey, H.~Lekkerkerker and G.~Maitland, \emph{Soft Matter},
  2007, \textbf{3}, 1145--1162\relax
\mciteBstWouldAddEndPuncttrue
\mciteSetBstMidEndSepPunct{\mcitedefaultmidpunct}
{\mcitedefaultendpunct}{\mcitedefaultseppunct}\relax
\EndOfBibitem
\bibitem[Divoux \emph{et~al.}(2013)Divoux, Grenard, and
  Manneville]{Divoux:2013}
T.~Divoux, V.~Grenard and S.~Manneville, \emph{Phys. Rev. Lett.}, 2013,
  \textbf{110}, 018304\relax
\mciteBstWouldAddEndPuncttrue
\mciteSetBstMidEndSepPunct{\mcitedefaultmidpunct}
{\mcitedefaultendpunct}{\mcitedefaultseppunct}\relax
\EndOfBibitem
\bibitem[Viasnoff and Lequeux(2002)]{Viasnoff:2002}
V.~Viasnoff and F.~Lequeux, \emph{Phys. Rev. Lett.}, 2002, \textbf{89},
  065701\relax
\mciteBstWouldAddEndPuncttrue
\mciteSetBstMidEndSepPunct{\mcitedefaultmidpunct}
{\mcitedefaultendpunct}{\mcitedefaultseppunct}\relax
\EndOfBibitem
\bibitem[Pignon \emph{et~al.}(1996)Pignon, Magnin, and Piau]{Pignon:1996}
F.~Pignon, A.~Magnin and J.-M. Piau, \emph{J. Rheol.}, 1996, \textbf{40},
  573--587\relax
\mciteBstWouldAddEndPuncttrue
\mciteSetBstMidEndSepPunct{\mcitedefaultmidpunct}
{\mcitedefaultendpunct}{\mcitedefaultseppunct}\relax
\EndOfBibitem
\bibitem[Coussot \emph{et~al.}(2002)Coussot, Nguyen, Huynh, and
  Bonn]{Coussot:2002c}
P.~Coussot, Q.~D. Nguyen, H.~T. Huynh and D.~Bonn, \emph{J. Rheol.}, 2002,
  \textbf{46}, 573--589\relax
\mciteBstWouldAddEndPuncttrue
\mciteSetBstMidEndSepPunct{\mcitedefaultmidpunct}
{\mcitedefaultendpunct}{\mcitedefaultseppunct}\relax
\EndOfBibitem
\bibitem[Coussot \emph{et~al.}(2006)Coussot, Tabuteau, Chateau, Tocquer, and
  Ovarlez]{Coussot:2006}
P.~Coussot, H.~Tabuteau, X.~Chateau, L.~Tocquer and G.~Ovarlez, \emph{J.
  Rheol.}, 2006, \textbf{50}, 975--994\relax
\mciteBstWouldAddEndPuncttrue
\mciteSetBstMidEndSepPunct{\mcitedefaultmidpunct}
{\mcitedefaultendpunct}{\mcitedefaultseppunct}\relax
\EndOfBibitem
\bibitem[Ovarlez and Coussot(2007)]{Ovarlez:2007}
G.~Ovarlez and P.~Coussot, \emph{Phys. Rev. E}, 2007, \textbf{76}, 011406\relax
\mciteBstWouldAddEndPuncttrue
\mciteSetBstMidEndSepPunct{\mcitedefaultmidpunct}
{\mcitedefaultendpunct}{\mcitedefaultseppunct}\relax
\EndOfBibitem
\bibitem[M{\o}ller \emph{et~al.}(2006)M{\o}ller, Mewis, and Bonn]{Moller:2006}
P.~C.~F. M{\o}ller, J.~Mewis and D.~Bonn, \emph{Soft Matter}, 2006, \textbf{2},
  274--283\relax
\mciteBstWouldAddEndPuncttrue
\mciteSetBstMidEndSepPunct{\mcitedefaultmidpunct}
{\mcitedefaultendpunct}{\mcitedefaultseppunct}\relax
\EndOfBibitem
\bibitem[Seth \emph{et~al.}(2012)Seth, Locatelli-Champagne, Monti, Bonnecaze,
  and Cloitre]{Seth:2012}
J.~Seth, C.~Locatelli-Champagne, F.~Monti, R.~Bonnecaze and M.~Cloitre,
  \emph{Soft Matter}, 2012, \textbf{8}, 140--148\relax
\mciteBstWouldAddEndPuncttrue
\mciteSetBstMidEndSepPunct{\mcitedefaultmidpunct}
{\mcitedefaultendpunct}{\mcitedefaultseppunct}\relax
\EndOfBibitem
\bibitem[Mansard \emph{et~al.}(2014)Mansard, Bocquet, and Colin]{Mansard:2014}
V.~Mansard, L.~Bocquet and A.~Colin, \emph{Soft Matter}, 2014, \textbf{10},
  6984--6989\relax
\mciteBstWouldAddEndPuncttrue
\mciteSetBstMidEndSepPunct{\mcitedefaultmidpunct}
{\mcitedefaultendpunct}{\mcitedefaultseppunct}\relax
\EndOfBibitem
\end{thebibliography}

\begin{thebibliography}{2}%
\providecommand*{\natexlab}[1]{#1}
\providecommand*{\mciteSetBstSublistMode}[1]{}
\providecommand*{\mciteSetBstMaxWidthForm}[2]{}
\providecommand*{\mciteBstWouldAddEndPuncttrue}
  {\def\EndOfBibitem{\unskip.}}
\providecommand*{\mciteBstWouldAddEndPunctfalse}
  {\let\EndOfBibitem\relax}
\providecommand*{\mciteSetBstMidEndSepPunct}[3]{}
\providecommand*{\mciteSetBstSublistLabelBeginEnd}[3]{}
\providecommand*{\EndOfBibitem}{}
\mciteSetBstSublistMode{f}
\mciteSetBstMaxWidthForm{subitem}
{(\emph{\alph{mcitesubitemcount}})}
\mciteSetBstSublistLabelBeginEnd{\mcitemaxwidthsubitemform\space}
{\relax}{\relax}


\bibitem[Mason \emph{et~al.}(1995)Mason, Bibette, and Weitz]{Mason:1995c}
T.~G. Mason, J.~Bibette and D.~A. Weitz, \emph{Phys. Rev. Lett.}, 1995,
  \textbf{75}, 2051--2054\relax
\mciteBstWouldAddEndPuncttrue
\mciteSetBstMidEndSepPunct{\mcitedefaultmidpunct}
{\mcitedefaultendpunct}{\mcitedefaultseppunct}\relax
\EndOfBibitem
\bibitem[Gallot \emph{et~al.}(2013)Gallot, Perge, Grenard, Fardin, Taberlet,
  and Manneville]{Gallot:2013}
T.~Gallot, C.~Perge, V.~Grenard, M.-A. Fardin, N.~Taberlet and S.~Manneville,
  \emph{Rev. Sci. Instrum.}, 2013, \textbf{84}, 045107\relax
\mciteBstWouldAddEndPuncttrue
\mciteSetBstMidEndSepPunct{\mcitedefaultmidpunct}
{\mcitedefaultendpunct}{\mcitedefaultseppunct}\relax
\EndOfBibitem
\end{thebibliography}

\clearpage
\onecolumngrid
\section*{\large{Supplemental material}}
\twocolumngrid
\appendix

\section{Supplemental movies}

\begin{figure}[!b]
\centering
 \includegraphics[width=0.9\linewidth]{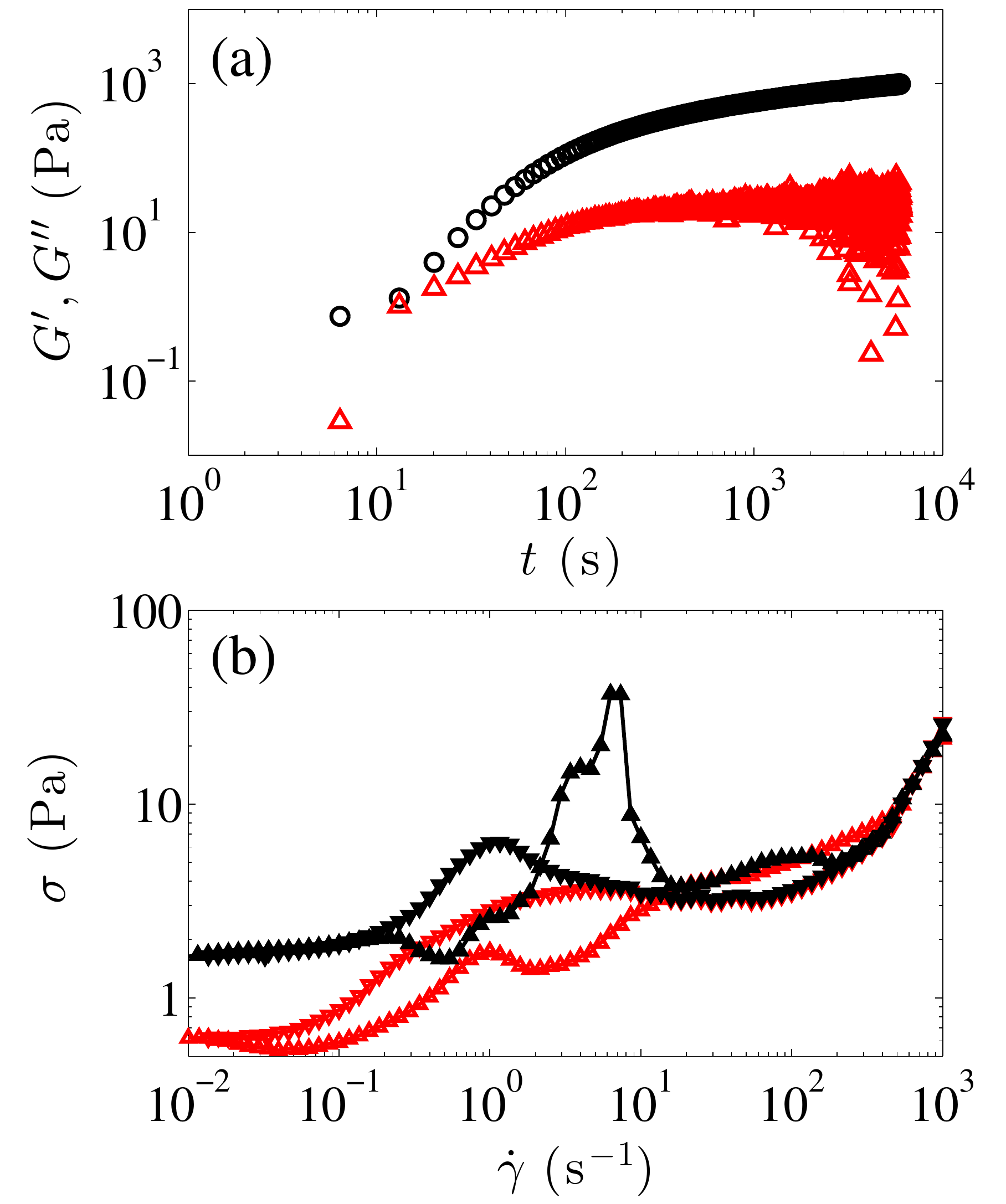}
  \caption{(a)~Elastic ($\circ$) and viscous moduli (\textcolor{red}{$\bigtriangleup$}) vs time after a preshear at $\dot \gamma_p=500$~s$^{-1}$ for 2~min ($f=1$~Hz, $\sigma=0.05$~Pa). Same data as in Fig.~2(a) in the main text. (b)~Flow curve, shear stress $\sigma$ $vs.$ applied shear rate $\dot\gamma$, obtained by decreasing ($\bigtriangledown$) and increasing ($\bigtriangleup$) shear rate in order from 10$^{-2}$ to 10$^{3}$ s$^{-1}$ with a waiting time of $8$~s per point. Black (resp. red) symbols correspond to results in a smooth (resp. rough) geometry.}
  \label{supfig1}
\end{figure}

Five movies are provided as supplemental materials to Figs.~5--9 of the main text. All movies illustrate the temporal evolution of the global rheology and of the 1D and/or 2D velocity profiles recorded simultaneously to the rheology during successive avalanche-like events.
 
\section{Supplemental figures}

Supplemental Fig.~\ref{supfig1}(a) displays the evolution of the elastic and viscous moduli of the sample after preshear. The gel rebuilds quickly as evidenced by the fact that the elastic modulus is larger than the viscous modulus about 20~s after preshear has been stopped.

Supplemental Fig.~\ref{supfig1}(b) shows flow curves $\sigma$ vs $\dot \gamma$ measured by rapidly sweeping down the shear rate from $10^{3}$ to $10^{-2}$~s$^{-1}$ and back up. The flow curves have complex non-monotonic shapes together with strong hysteresis under both smooth and rough boundary conditions.

\begin{figure}[!b]
\centering
 \includegraphics[width=0.9\linewidth]{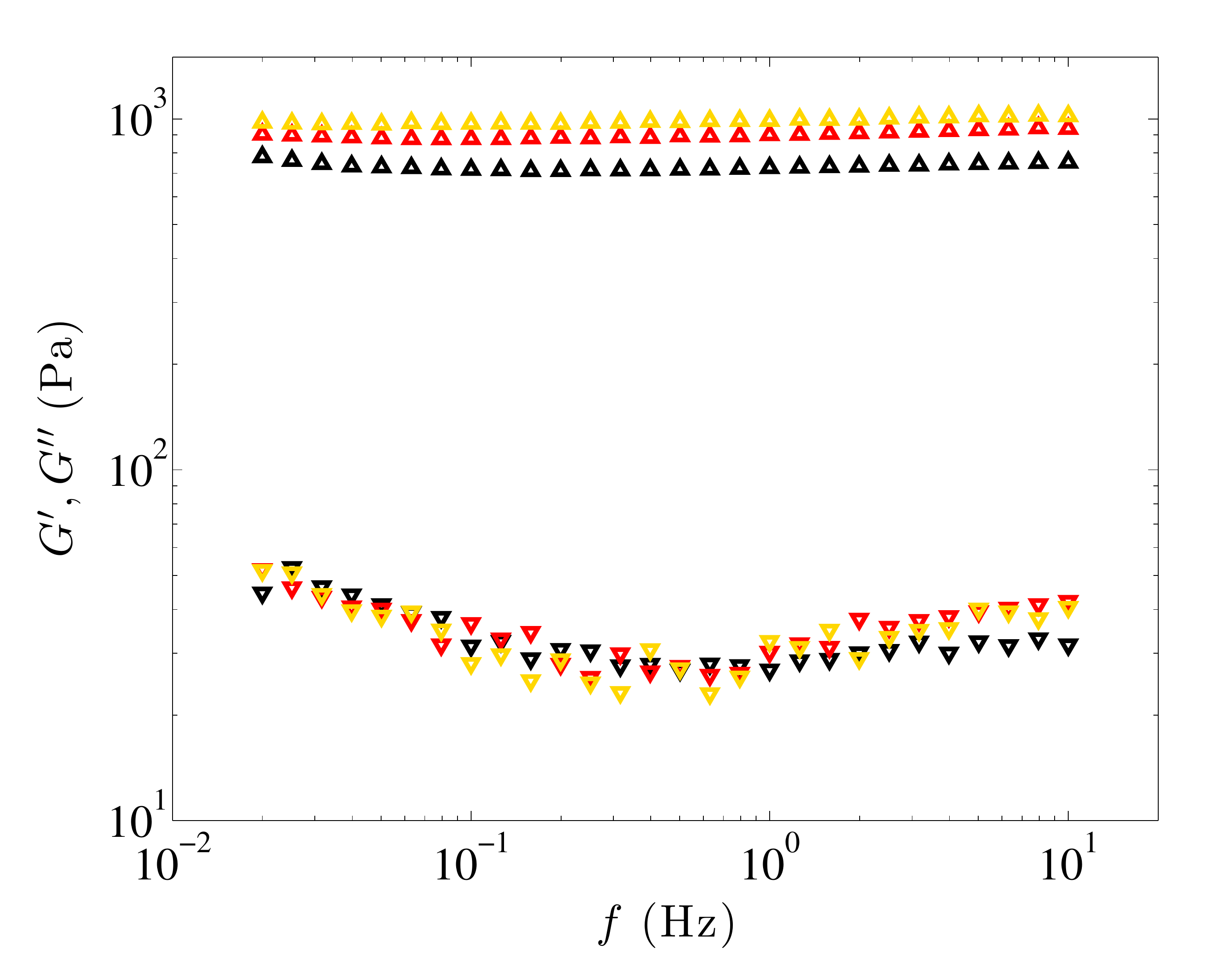}
  \caption{Elastic ($\bigtriangleup$) and viscous ($\bigtriangledown$) moduli vs frequency $f$. The frequency sweep is performed at a constant strain $\gamma=0.5$~\% and at different aging times $t_w$ after preshear. [color, $t_w$ (min)]: (\textcolor{black}{\Large{-}},~30); (\textcolor{red}{\Large{-}},~60); (\textcolor{yellow}{\Large{-}}, 100). }
  \label{supfig2}
\end{figure}

Supplemental Fig.~\ref{supfig2} illustrates the result of a frequency sweep performed at a constant strain ($\gamma=0.5$~\%) from 10 to 0.02~Hz, and for different aging times $t_w=30$, 60 and 100~min after stopping the preshear. The elastic modulus $G'$ is independent of the frequency and increases with the aging time $t_w$ in agreement with Supplemental Fig.~\ref{supfig1}. The viscous modulus shows a broad minimum with no sign of increasing part at low frequencies, a feature that is typical of soft glassy systems  \cite{Mason:1995c}. 

\begin{figure}[!t]
\centering
 \includegraphics[width=0.9\linewidth]{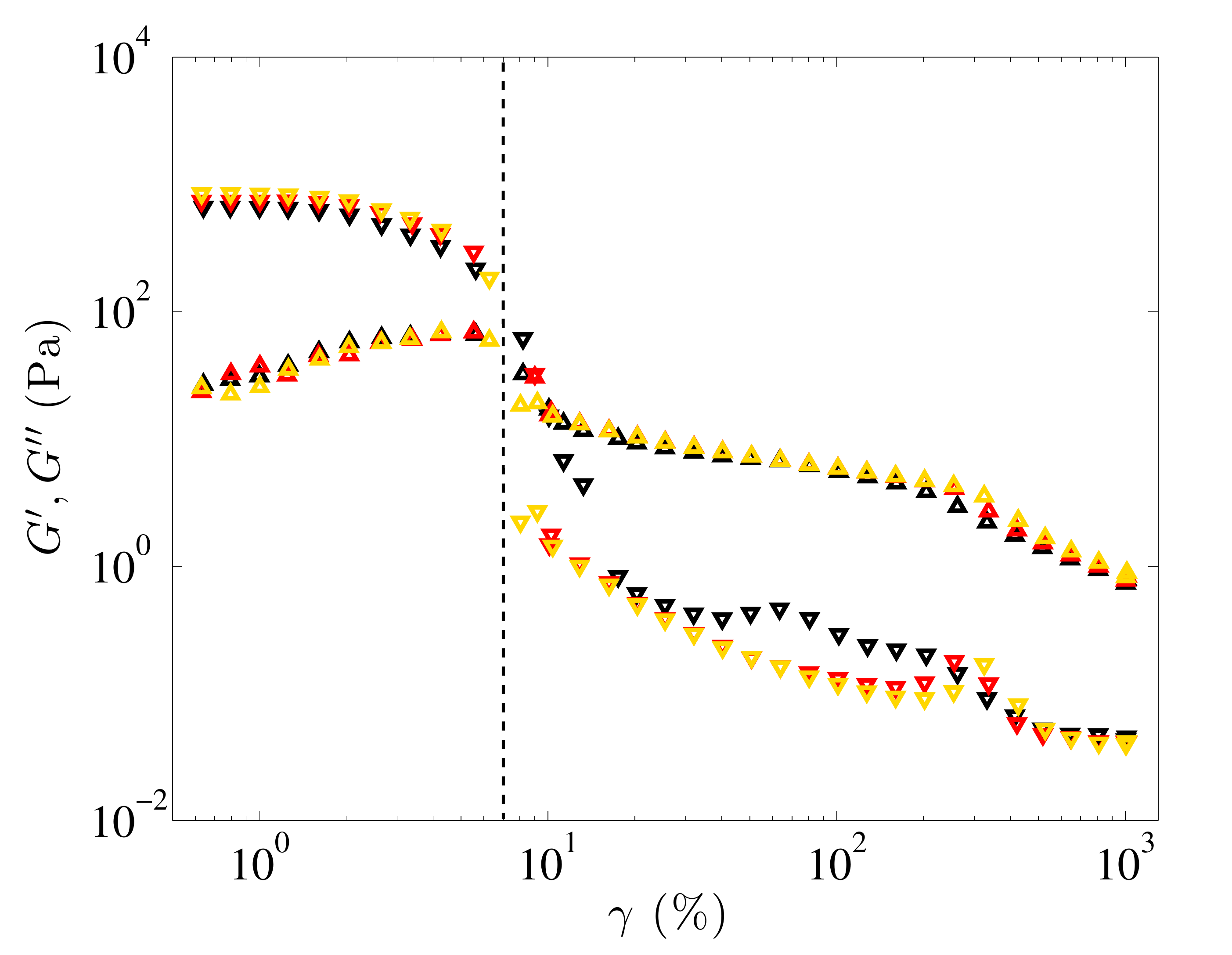}
  \caption{Elastic ($\circ$) and viscous ($\bigtriangleup $) moduli vs the strain amplitude. The strain sweep is performed at a fixed frequency $f=1$~Hz with a waiting time of 8~s per point, and at different times $t_w$ after the preshear. [color, $t_w$ (min)]: (\textcolor{black}{\Large{-}},~30); (\textcolor{red}{\Large{-}},~60); (\textcolor{yellow}{\Large{-}}, 100). The vertical dashed line at $\gamma \sim 7$~\% emphasizes the yield point defined as the point at which $G'=G''$.}
  \label{supfig3}
\end{figure}

Supplemental Fig.~\ref{supfig3} shows the results of a strain sweep performed on the gel at a constant frequency $f=1$~Hz for different aging times $t_w$ after preshear. The gel yields at a constant strain of about 7~\% that is roughly independent of the aging time $t_w$.  

\begin{figure}[t!]
\centering
 \includegraphics[width=\linewidth]{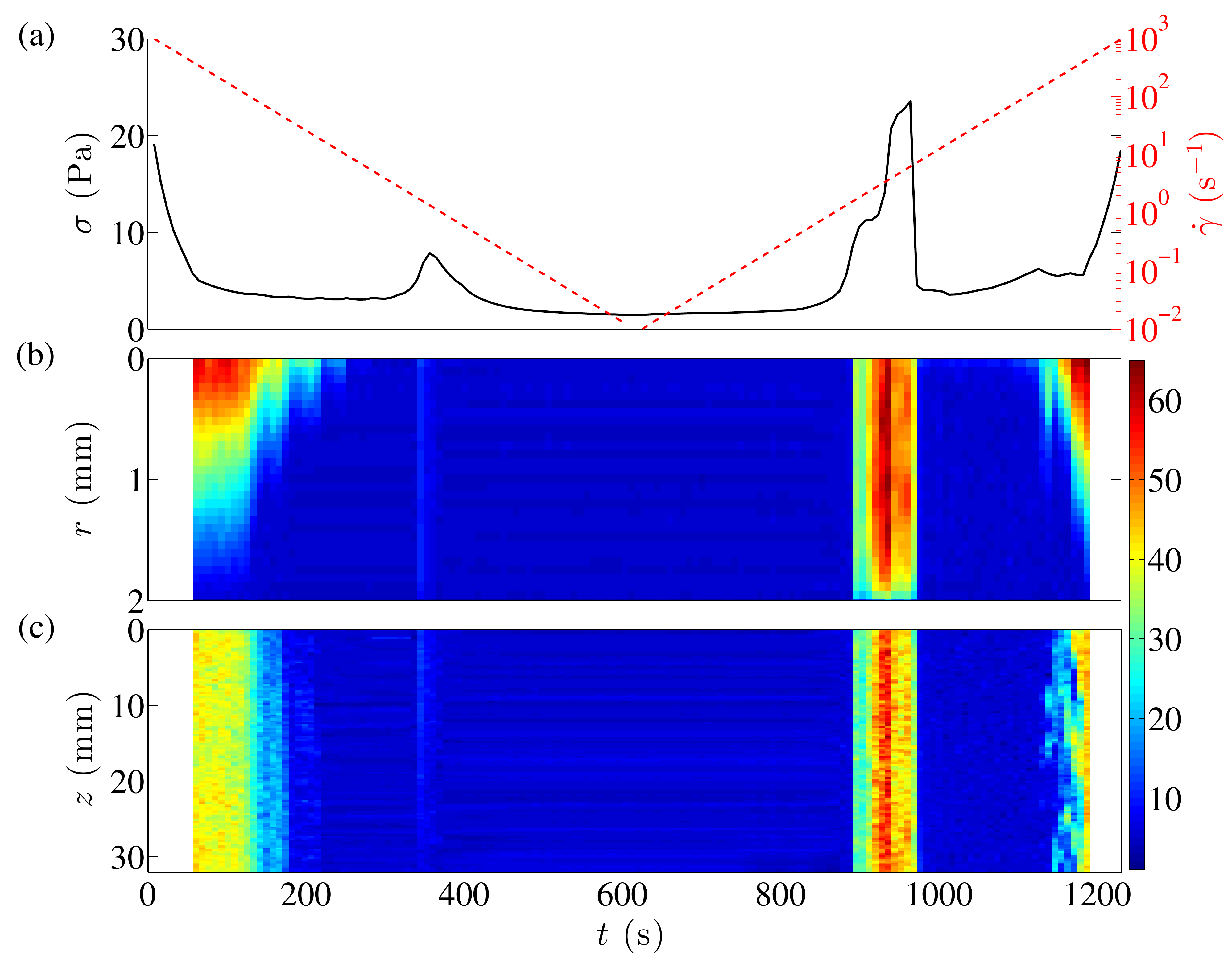}
  \caption{Taylor-Couette cell with smooth boundary conditions. (a) Shear rate $\dot \gamma$ and shear stress  $\sigma$ vs time $t$ obtained by first decreasing continuously $\dot \gamma$ from $10^3$ to $10^{-2}$~s$^{-1}$ in 75 logarithmically spaced points of duration $\delta t=8$~s each, and then increasing $\dot \gamma$ over the same range. (b)~Spatiotemporal diagram of the velocity data $v(r,z_0,t)$ as a function of the distance $r$ to the rotor and time $t$, at $z_0\simeq 15$~mm. (c)~Spatiotemporal diagram of the velocity data $v(r_0,z,t)$ as a function of the vertical position $z$ and time $t$, at $r_0=0.5$~mm. The vertical position $z$ is measured from the upper boundary of the 2D-USV probe. The fluid velocity is color-coded in mm.s$^{-1}$.}
  \label{supfig4}
\end{figure}

\begin{figure}[!t]
\centering
 \includegraphics[width=\linewidth]{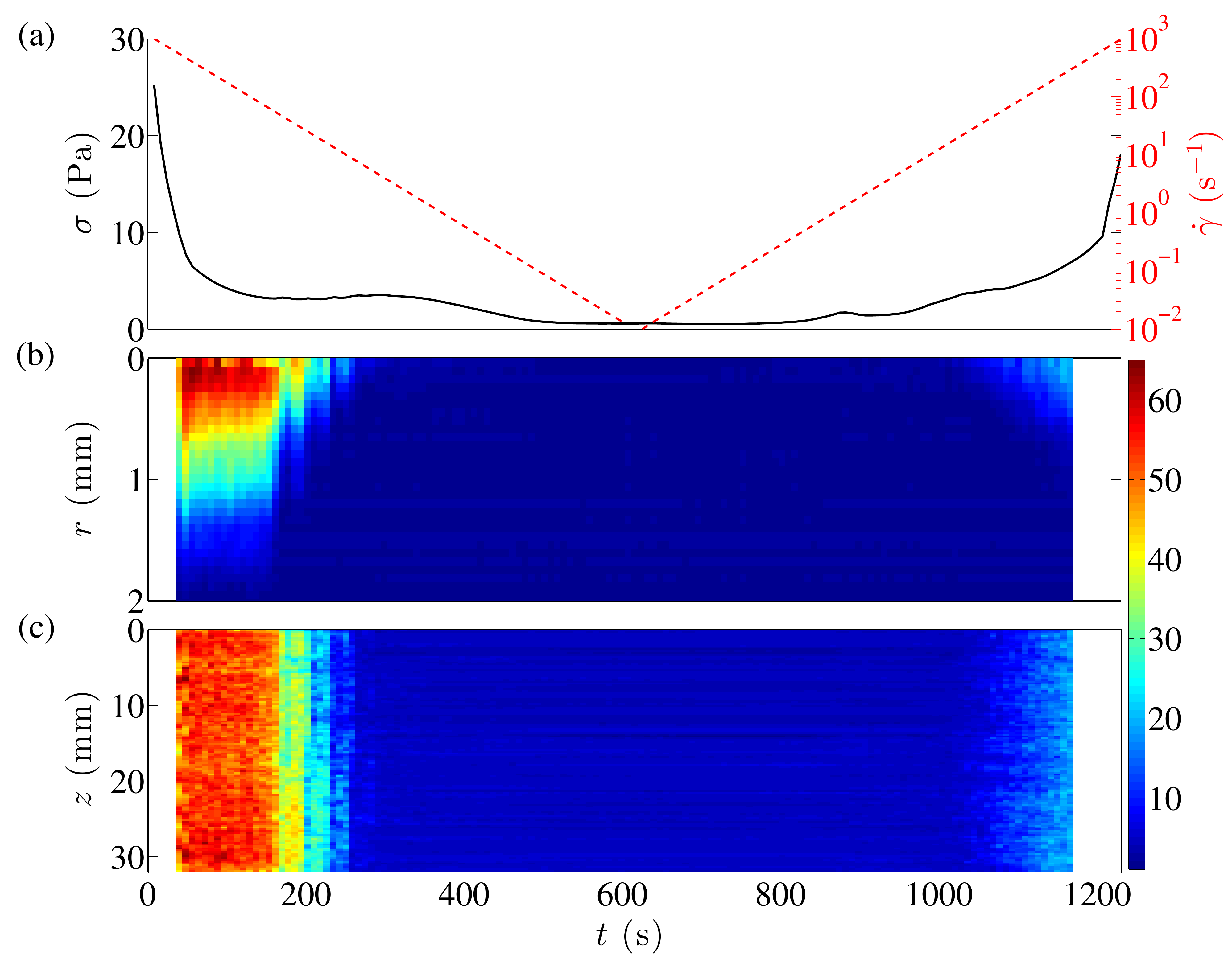}
  \caption{Same as Supplemental Fig.~\ref{supfig4} for a Taylor-Couette cell with rough boundary conditions.}
  \label{supfig5}
\end{figure}

Supplemental Figs.~\ref{supfig4} and \ref{supfig5} show the rheological and 2D-USV data obtained in smooth and rough boundary conditions respectively, by first decreasing the shear rate and then increasing it back again. The corresponding flow curves are shown in Fig.~1(b) of the present ESI. Here we also show spatiotemporal plots of the velocity profiles as a function of the radial position $r$ at a given height ($z_0\simeq 15$~mm) in the Taylor-Couette cell [Supplemental Figs.~\ref{supfig4}(b) and \ref{supfig5}(b)], and as a function of the vertical position $z$, at a given radial position inside the gap ($r_0=0.5$~mm) [Supplemental Figs.~\ref{supfig4}(c) and \ref{supfig5}(c)]. Velocity data are shown only for $\dot \gamma <$460 and 340~s$^{-1}$ in smooth and rough boundary conditions respectively, due to the limitation in the ultrasonic pulse repetition frequency to 20~kHz, which sets an upper bound on the measurable velocities \cite{Gallot:2013}. 

For smooth boundary conditions, the flow curve exhibits a large hysteresis, mainly localized in a narrow range of shear rates [Supplemental Fig.~\ref{supfig1}(b)]. During the decreasing ramp of $\dot \gamma$, the sample which is first sheared homogeneously at large shear rates, enters a total wall slip regime at $t \simeq 200$~s, i.e., for $\dot \gamma \sim 20$~s$^{-1}$ [Supplemental Fig.~\ref{supfig4}(b)]. This regime persists until $\dot \gamma$ is increased back up again. The subsequent fluidization of the material is abrupt and occurs in a narrow range of shear rates, $1 \lesssim \dot \gamma \lesssim 10$~s$^{-1}$, which coincides with the brutal increase of the global stress. Supplemental Fig.~\ref{supfig4}(b,c) reveal that the subsequent drop in the stress corresponds to flow arrest, i.e. total wall slip, until the sample fluidizes again for $\dot \gamma \gtrsim 100$~s$^{-1}$ and that the local behavior of the sample is roughly homogeneous along the whole height of the cell during both ramps. 

\begin{figure*}[t!]
\centering
 \includegraphics[width=0.7\linewidth]{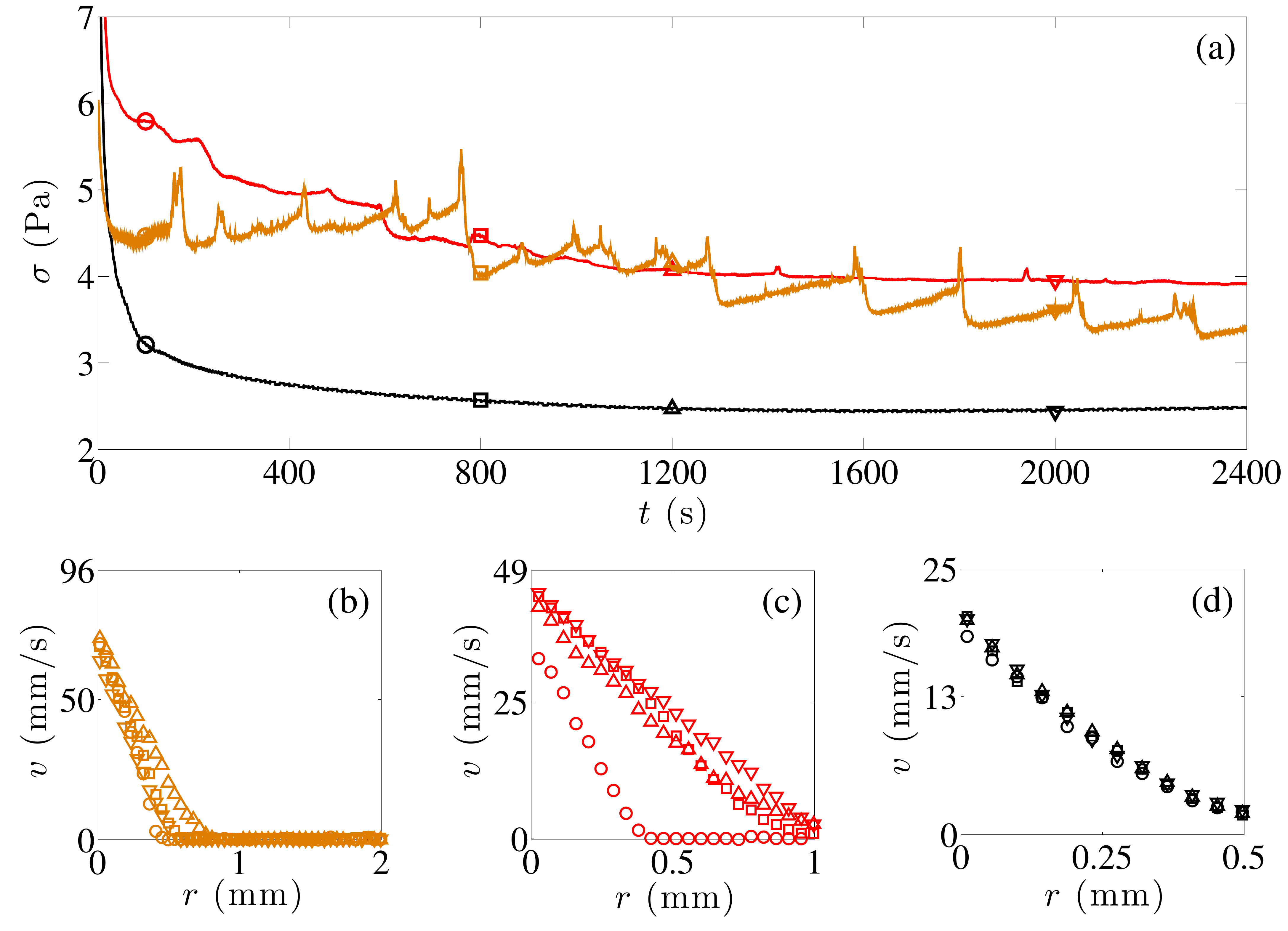}
  \caption{(a) Stress response $\sigma(t)$ to a shear startup experiment performed at $\dot \gamma=50$~s$^{-1}$ for different gap sizes: [color, gap (mm)] = [\textcolor{brown}{\textbf{\large --}}, 2], [\textcolor{red}{\textbf{\large --}},~1], [\textcolor{black}{\textbf{\large --}}, 0.5]. (b)-(d) Velocity profile, where $r$ is the distance to the rotor at different times in (a) [symbol, time (s)]: [$\bigcirc$, 100]; [$\Box$, 800]; [$\bigtriangleup$, 1200]; [$\bigtriangledown$, 2000]. The rotor velocity corresponds to the upper bound of the vertical axis. The sample is aged during $t_w=10$~min before each experiment.
}
  \label{supfig6}
\end{figure*}

Supplemental Fig.~\ref{supfig5} illustrates the exact same protocol as that reported in Supplemental Fig.~\ref{supfig4}, but performed in a rough Taylor-Couette cell. As also seen in Supplemental Fig.~\ref{supfig1}(b), the rheological hysteresis is far less pronounced in comparison to the data obtained with smooth boundary conditions. Here the flow remains homogeneous along the vertical axis during both ramps [Supplemental Fig.~\ref{supfig5}(c)]. The velocity profiles at a given height in the Taylor-Couette cell also display very different behavior from the smooth case. In the decreasing ramp of $\dot \gamma$, the stress plateaus in between $\dot \gamma \sim 100$~s$^{-1}$ and $\dot \gamma \sim 1$~s$^{-1}$, while shear localizes at the rotor. Below $\dot \gamma \sim 1$~s$^{-1}$, the stress displays a kink towards a second plateau, while the sample exhibits a plug flow. Increasing $\dot \gamma$ back up, the plug flow persists up to shear rates of a few 10~s$^{-1}$. Above the latter value, the fluidization of the sample proceeds from the rotor and involves transient banding. Full fluidization is not available in Supplemental Fig.~\ref{supfig5}(b) for the technical reason mentioned above. 

Supplemental Fig.~\ref{supfig6} shows the influence of the gap size on the fluid response to a shear startup experiment performed at $\dot \gamma=50$~s$^{-1}$ on a sample left at rest for $t_w=10$~min. In a gap of size $e=2$~mm, the same value as in the main text, the stress response displays a large number a peaks together with unsteady shear banding over 2400~s [Supplemental Fig.~\ref{supfig6}(b)]. Decreasing the size of the gap changes radically the sample behavior. In a gap twice as small ($e=1$~mm), the stress response exhibits less peaks and the sample fluidizes entirely in about $t=200$~s [Supplemental Fig.~\ref{supfig6}(c)]. Decreasing the gap by a factor of 4 ($e=0.5$~mm) leads to a smooth stress response and to homogeneous velocity profiles from the beginning of the experiments, with significant slip at the moving boundary [Supplemental Fig.~\ref{supfig6}(d)].

 \end{document}